\documentclass[preprint,nofootinbib,iopjournal,superscriptaddress,eqsecnum]{revtex4-2}
\usepackage[top=100pt,bottom=0pt,letterpaper]{geometry}
\usepackage{adjustbox}
\usepackage{natbib}
\usepackage[T1]{fontenc}
\usepackage{lmodern}
\usepackage[utf8]{inputenc}
\usepackage{microtype}
\usepackage{comment}
\usepackage{xcolor}
\usepackage[normalem]{ulem}
\usepackage{setspace}
\usepackage{float}
\restylefloat{table}

\usepackage{appendix}
\usepackage{dcolumn}
\usepackage{lipsum}
\usepackage{graphicx,epsfig}
\usepackage{psfrag}
\usepackage{bm}
\usepackage{epstopdf}
\usepackage{latexsym}
\usepackage{multirow}
\usepackage{amsmath,amssymb}
\usepackage{enumerate}
\usepackage{hyperref}
\usepackage[FIGTOPCAP]{subfigure}
\usepackage{empheq,mathtools}
\usepackage{booktabs}
\usepackage{tabularx}
\usepackage{longtable}
\def\be {\begin{equation}}
\def\ee {\end{equation}}
\def\ba {\begin{eqnarray}}
\def\ea {\end{eqnarray}}
\def\nn {\nonumber}
\def\bc {\begin{center}}
\def\ec {\end{center}}
\newcommand{\bdm}{\begin{displaymath}}
\newcommand{\edm}{\end{displaymath}}

\def\a  {\alpha}
\def\b  {\beta}

\def\G  {\Gamma}
\def\d  {\delta}
\def\D  {\Delta}
\def\ep  {\epsilon}

\def\l  {\lambda}
\def\m  {\mu}

\def\r  {\rho}
\def\th {\theta}

\def\t  {\tau}

\def\z {\zeta}

\def\mc {\mathcal}
\def\nn {\nonumber}

\def\ra {\rightarrow}

\def\la {\label}
\def\le {\left}
\def\ri {\right}

\def\f {\frac}
\def\sq {\sqrt}
\def\bi {\begin{itemize}}
\def\ei {\end{itemize}}

\def\> {\rangle}
\def\< {\langle}

\def\bc {\begin{center}}
\def\ec {\end{center}}

\usepackage{amsfonts}

\begin{document}
\title{Gravitational lensing and deflection angles of generalised Ellis-Bronnikov wormhole embedded in a warped braneworld background}

\author{Soumya Jana}
\email[Email address: ]{soumyajana.physics@gmail.com}
\affiliation{Department of Physics, Sitananda College, Nandigram, 721631, India}
\affiliation{Department of Physics, Indian Institute of Technology, Kharagpur 721 302, India}

\author{Vivek Sharma} \email[email: ]{svivek829@gmail.com}
\affiliation{Dr. Bhawar Singh Porte Govt. PG College, Chhattisgarh - 495119, India}

\author{Suman Ghosh} \email[email: ]{suman.ghosh@bitmesra.ac.in (corresponding author)}
\affiliation{Department of Physics, Birla Institute of Technology, Ranchi - 835215, India}

\begin{abstract}

\noindent We investigate null geodesics, deflection angles, and gravitational lensing by the generalized Ellis–Bronnikov (GEB) wormhole and its embedding in a five-dimensional warped braneworld (WGEB). The GEB geometry extends the standard Ellis–Bronnikov wormhole through a steepness parameter $m\geq 2$, which controls the throat profile while partially reducing the violation of the classical energy conditions. We derive analytic expressions for the weak- and strong-deflection angles and obtain the corresponding Einstein-ring and relativistic-image relations. The standard Ellis–Bronnikov results are recovered as the special case $m=2$, while for $m>2$ the strong-deflection angle exhibits a power-law divergence instead of the familiar logarithmic behaviour. For the warped five-dimensional geometry, we show that the extra-dimensional momentum modifies the effective impact parameter, and hence the deflection angle and image positions while leaving the location of the unstable photon orbit unchanged. We further derive generalized relations between relativistic and non-relativistic Einstein-ring radii. 
Our analysis shows that deflection/ring data alone cannot distinguish the effect of the warped extra dimension in WGEB scenario.

\end{abstract}
\maketitle

\newpage
\section{introduction}
Wormholes create ``short-cuts'' that enable ``apparently faster than light'' movement between two distant spacetime sites \cite{Visser:1995cc,Alcubierre:2017pqm,Witten:2019qhl}. 
The earliest example, the Einstein–Rosen bridge \cite{Einstein:1935tc}, was initially proposed as a bridge-like connection between two asymptotically flat spacetimes. However, it was soon realized that such a structure is non-traversable, since it collapses too rapidly for even light signals to pass through it \cite{Fuller:1962zza}.
The concept of traversable wormholes was later developed systematically by Morris and Thorne \cite{PhysRevLett.61.1446,Morris:1988cz}, who showed that stable traversable wormholes require the violation of classical energy conditions, particularly the null energy condition (NEC).

The requirement of exotic matter with negative energy density remains one of the major obstacles to the physical realization of wormholes within classical general relativity \cite{Alcubierre:2017pqm, Witten:2019qhl, Roman:2004xm, Hochberg:1998ha}.
Although quantum effects may permit the formation of microscopic wormholes \cite{Gao:2016bin,Maldacena:2018gjk,Fu:2019vco}, such mechanisms are generally insufficient to support macroscopic traversable configurations. This difficulty has motivated extensive studies of wormholes in modified theories of gravity, where effective violations of energy conditions may arise from geometric corrections rather than from exotic matter sources \cite{Hochberg:1990is, Bhawal:1992sz, Agnese:1995kd, Samanta:2018hbw, Lobo:2008zu, Kanti:2011jz, Kanti:2011yv, Zubair:2017oir, Shaikh:2016dpl, Ovgun:2018xys, Canate:2019spb}. Dynamical wormhole models \cite{Hochberg:1998ii, Roman:1992xj, Kar:1994tz, Kar:1995ss, Visser:2003yf}, higher-curvature theories such as f(R) gravity \cite{Lobo:2009ip, Garcia:2010xb, MontelongoGarcia:2010xd, Sajadi:2011oei}, Born–Infeld gravity \cite{Maeda:2008nz,Shaikh:2015oha,Shaikh:2018yku}, and torsional gravity models  \cite{Boehmer:2012uyw, Bronnikov:2015pha, DiGrezia:2017daq} have all provided viable frameworks in which traversable wormholes may exist while reducing or avoiding explicit NEC violation.

A central question is whether such wormholes, if they exist in nature, can be distinguished observationally from black holes or other compact objects. Gravitational lensing provides one of the most promising avenues for such tests \cite{Cramer:1994qj, Safonova:2001vz, Bozza:2002zj, Dey:2008kn,Shaikh:2018oul}. Since wormholes can act as gravitational lenses, the resulting image positions, Einstein rings, relativistic images, and photon spheres may carry characteristic signatures of the underlying geometry. Recent advances in gravitational-wave astronomy \cite{Bishop:2021rye, Arimoto:2021cwc, Dent:2020nfa}, black hole imaging \cite{EHT2019_I, Johnson2023ngEHT, Kardashev2013}, and the possibility of astrophysical wormholes in galactic dark matter halos \cite{rahaman2014possible, Rahaman:2013xoa} have further strengthened interest in identifying observable wormhole signatures. In addition, wormholes may behave as black-hole mimickers, leaving distinctive imprints in quasi-normal modes, ringdown signals, and compact-object mergers \cite{Krishnendu:2017shb, Cardoso:2016oxy, Aneesh:2018hlp, DuttaRoy:2019hij, Mitra:2023yjf}.

On a different but closely related front, theories with extra spatial dimensions have long played an important role in attempts to unify fundamental interactions and address hierarchy problems in high-energy physics \cite{Kaluza:1921tu, Klein:1926tv, Green:1987sp, Rubakov:1983bb,Gogberashvili:1998vx}. From the original Kaluza–Klein framework \cite{Kaluza:1921tu, Klein:1926tv} to superstring theory \cite{Green:1987sp} and modern braneworld scenarios \cite{Rubakov:1983bb,Gogberashvili:1998vx}, higher-dimensional models arise naturally from fundamental symmetry principles rather than as ad hoc extensions of general relativity. Among them, warped braneworld models  \cite{Gogberashvili:1998iu, Randall:1999ee, Randall:1999vf} are particularly important because they introduce a non-factorizable geometry in which the four-dimensional spacetime is embedded in a curved higher-dimensional bulk. The extra dimension influences the effective four-dimensional geometry through a warp factor, leading to potentially observable deviations from standard gravitational behavior.
Five dimensional braneworld Wormhole models have also been constructed and studied in \cite{Lobo:2007qi,deLeon:2009pu, Wong:2011pt, Kar:2015lma, Banerjee:2019ssy, Wang:2017cnd, Kar:2022omn}.

The generalised Ellis–Bronnikov (GEB) family, introduced in \cite{Kar:1995jz}, generalises the standard Ellis–Bronnikov (EB) wormhole \cite{Ellis:1973yv, Bronnikov:1973fh} by introducing a steepness parameter $m$, which controls the geometry of the throat while reducing the violation of energy conditions. 
Possible matter source of GEB wormholes has also been discussed in \cite{Crispim:2024dgd}. 
GEB models have recently been constructed within the framework of asymptotically safe gravity \cite{Nilton:2022hrp}, $f(R)$ gravity in \cite{Sokoliuk:2022xcf} and scalar-Einstein-Gauss-Bonnet \cite{Ernazarov:2025fyv}. 
Recently, the embedding of four-dimensional generalized Ellis–Bronnikov (GEB) wormhole in a static five-dimensional warped braneworld background (WGEB) has been constructed in \cite{Sharma:2021kqb} and was shown to further reduce the violation of energy conditions, particularly in the presence of a decaying warp factor. \footnote{Such 5D geometry with a cosmological thick brane in the presence of both growing and decaying warp factor has been constructed in \cite{Ghosh:2008vc}.}
Earlier studies have analysed the geometry, timelike geodesics, geodesic congruences, and quasi-normal modes (QNMs) of these spacetimes \cite{Sharma:2022tiv, Sharma:2022dbx, Mitra:2023yjf}, revealing signatures of the extra dimension through confinement properties and multiple quasi-normal ringing phases.
Recently, dyonic Ellis-Bronnikov wormhole is constructed, within a four-dimensional effective theory derived from a five-dimensional theory with a warped Kaluza-Klein compactification, in \cite{Lobo:2026eoh}.
While several aspects of gravitational lensing for the standard EB wormhole have been studied previously \cite{Bhattacharya:2010zzb, Tsukamoto:2012xs, Tsukamoto:2012zz, Nakajima:2012pu, Tsukamoto:2016zdu, Tsukamoto:2017edq, Tsukamoto:2017hva, Tsukamoto:2022uoz, Ahmed:2025vww}, a systematic analysis for the generalized GEB family and its warped braneworld embedding has not been carried out so far (see Table \ref{tab:lit} for a summary of previous work and comparison with the present work). 
Since gravitational lensing directly probes null geodesics and provides one of the most promising observational tests of compact objects, it is natural to ask how the generalized throat geometry and the extra-dimensional momentum modify the deflection angle, Einstein-ring structure and relativistic images. Addressing these questions constitutes the primary objective of the present work.

\begin{table}
    \centering
    \begin{tabular}{|l|c|c|}
    \hline
        Geometry  & References  & Scope / Main focus \\
          \hline
       EB  &  \cite{Bhattacharya:2010zzb, Tsukamoto:2012xs, Tsukamoto:2012zz, Nakajima:2012pu, Tsukamoto:2016zdu, Tsukamoto:2017edq, Tsukamoto:2017hva, Tsukamoto:2022uoz, Ahmed:2025vww} &  Weak and Strong Lensing, Einstein Rings \\
       \hline
       GEB  & \cite{Kar:1995jz, Sokoliuk:2022xcf, Ernazarov:2025fyv,Nilton:2022hrp, Crispim:2024dgd, DuttaRoy:2019hij, Mitra:2023yjf}  & Model building, geometry, QNM \\
          \hline
       5D-GEB      & \cite{Kar:2015lma, Sengupta:2021wvi, Kar:2022omn, Sharma:2021kqb, Sharma:2022dbx, Sharma:2022tiv, Mitra:2023yjf, Lobo:2026eoh} & Model building, geometric features  \\     \hline
        GEB,  WGEB  &  Present work & Weak and Strong Lensing, Einstein Rings \\
             \hline
    \end{tabular}
    \caption{Summary of the scope of earlier works on gravitational lensing by EB/GEB/5D-GEB wormholes in comparison with the focus of the present work.}
    \label{tab:lit}
\end{table} 

Our goal is to identify potential observable signatures of the steepness parameter $m$, the throat radius $b_0$ and the extra-dimensional (dimensionless) parameter $\d$, which represents the relative momentum of photons along the warped extra dimension.
We derive analytic expressions for the deflection angle and lensing relations in both weak- and strong-lensing regimes and recover the known Ellis–Bronnikov results in the limit $m=2$. For general $m\ge2$, numerical analysis is used alongside analytic approximations. We demonstrate that the parameter $m$ leaves clear imprints on lensing observables such as deflection angles, Einstein ring radii, and image positions, while the effect of the extra dimension is to modify the effective impact parameter and consequently alters the lensing observables.
Although modified lensing signatures are not unique to WGEB scenario and may also arise in other modified theories of gravity, the deviations arise in the present model from the conserved momentum associated with the extra dimension.

The paper is organized as follows. In Sec. \ref{sec:geo}, we briefly review the GEB and WGEB wormhole geometries. In Sec. \ref{sec:Null-Trajectories}, we analyse null trajectories and classify the possible photon orbits. In Sec. \ref{sec:def}, we derive the deflection angles in weak and strong-lensing limits. Gravitational lensing and Einstein ring relations are studied in Sec. \ref{sec:lensing}. The distinguishability of wormhole parameters and the extra dimension are discussed in Sec. \ref{sec:obs}. Our findings are summarized systematically in Sec. \ref{sec:discussion}.


\section{Wormhole Geometry} \la{sec:geo}

We begin by briefly reviewing the five-dimensional warped geometry in which the four-dimensional generalized Ellis–Bronnikov wormhole is embedded. The general form of such a warped spacetime is given by
\begin{equation}
ds^{2} = g_{\mu \nu}~dx^{\mu}~dx^{\nu} + g_{44}~(dx^{4})^{2} ,\label{eq:5d-general-line-element}
\end{equation}
where, $g_{\mu \nu}~dx^{\mu}~dx^{\nu}$  represents the 4 dimensional part  ($\mu$ runs from $0$ to $3$  and $x^4$ represents the extra dimension). In the so-called tortoise coordinate $l \in{(-\infty, \infty} $), the WGEB model \cite{Sharma:2021kqb} is given by
\begin{equation}
ds^{2} =  e^{2f(y)} \Big[ - dt^{2} +  dl^{2} + r^{2}(l)~\big(  d\theta^{2} + \sin^{2}(\theta)~d\phi^{2} \big) \Big] + dy^{2} , \label{eq:5d-line-element}
\end{equation}
where, $y$ is the extra dimension ($ - \infty \leq y \leq \infty$), $f(y)$ is a warping factor. Throughout this article, we choose $f(y) = \pm \log[\cosh(y/y_{0})]$ which corresponds to a thick  brane scenario \cite{Dzhunushaliev:2009va}. The plus and minus signs imply growing and decaying warp factor respectively. Such analytic expressions of the warping factor were found as solutions of Einstein equations in presence of tachyon and phantom scalar field, respectively in \cite{Koley:2004at,Zhang:2007ii}. Then, at $y = 0$, metric (\ref{eq:5d-line-element}) reduces to a GEB geometry given by \cite{Kar:1995jz}, 
\begin{equation}
ds_4^{2} = - dt^{2} +  dl^{2} + r^{2}(l)~\big(  d\theta^{2} + \sin^{2}(\theta)~d\phi^{2} \big) \label{eq:generalised-E&B}
\end{equation}
where, \begin{equation}
r(l) = (b_{0}^{m} + l^{m})^{1/m} , ~~~~ m = 2, 4, 6 ....\label{eq:r(l)}
\end{equation}
Metric (\ref{eq:generalised-E&B}) is spherically symmetric and ultra-static model in which $b_{0}$ represents the neck or throat radius of wormhole. 
The parameter $m$ controls the steepness of the wormhole throat. Larger values of $m$ correspond to sharper flattening of the geometry away from the throat.
Note that the function $r(l)$ is smooth in the entire domain of proper radial distance, $l$ ($-\infty \leq l \leq \infty$).
For odd integer values of $m$, the radial coordinate $r(l)$ becomes negative for $l < -b_0$, which is incompatible with its geometric interpretation as the radius of a two-sphere.
In addition, non-integer values of $m$ result in complex values of $r(l)$.
The restriction on $m$ to be positive even integers, therefore guarantees a smooth wormhole geometry.
In terms of the radial coordinate, Eq. (\ref{eq:generalised-E&B}) can be written as,
\begin{equation}
ds^{2} = -dt^{2} + \frac{dr^{2}}{ 1 - \frac{b \left( r \right)}{r}} + r^{2} \left( d\theta^{2} + \sin^{2} \theta d\phi^{2} \right) \label{eq:geb_metric_r} 
\end{equation}
Where $r$ and $l$ are related through the shape function $b(r)$ as,
\begin{equation}
dl^{2} = \frac{dr^{2}}{\left( 1 - \frac{b(r)}{r} \right)} \Longrightarrow b(r) = r - r^{\left( 3 - 2m\right)} \left( r^{m} - b_{0}^{m} \right)^{\left( 2 - \frac{2}{m} \right)} .\label{eq:r_l_relation}
\end{equation}
One can get back the EB space-time for $m = 2$ which is a static, spherically symmetric, geodesically complete, horizonless space-time constructed by using phantom scalar field (a scalar field with negative kinetic term in stress-energy-tensor violating null energy condition). 
The Ricci scalar for the 4D geometry is given by 
\be
{}^{(4)}R = -\f{2}{r^2} \le( r'^2 + 2rr'' -1 \ri). \la{eq:ricci-4D}
\ee
Note that, about $l \sim 0.63~ b_0$ distance away from the neck, the curvature falls below $1/b_0^2$. For higher $m$ -geometries the curvature begins to flatten out at even smaller distance from the neck. This implies that, trajectories that pass by at even further distance will feel lesser effect of the wormhole. Therefore, the characteristic size of the wormhole may be effectively identified with the throat radius $b_0$.

It was shown in \cite{Sharma:2021kqb, Mitra:2023yjf} that for WGEB models two of the (on-brane) weak energy conditions, $\r > 0$ and $\r + p > 0$, are satisfied in the presence of a decaying warp factor. 
The weak energy condition violation comes from the radial pressure $\t$ as
\be
\r(l,y) + \t(l,y) = -2 e^{-2f} \f{r''}{r} \leq 0 ,  \la{eq:wec-t}
\ee
which is non-positive everywhere and vanishes at the throat for $m >2$. 
However, for a decaying warp factor, the pre-factor in Eq. (\ref{eq:wec-t}) has its minima at the location of the brane.
In thick brane models, it has been shown that \cite{Dzhunushaliev:2009va}, a warped geometry with growing warp factor is sourced by negative energy bulk tachyon fields, whereas a decaying warp factor shows up in the presence of matter fields with positive energy density, e.g. a scalar field with Sine-Gordon potential. Thus in presence of a decaying warp factor, the negative energy density (needed for the 4D wormhole geometry) is partially compensated for. 
This results in reduced violation of the energy conditions.
This further motivated the study of geometry, curvature invariants, geodesics and geodesic-congruences in WGEB spacetimes  \citep{Sharma:2021kqb, Sharma:2022tiv, Sharma:2022dbx}. More recently, the stability analysis or the quasi-normal modes of the WGEB geometry is studied that shows signature of the extra dimension through two distinct quasi-normal ringing era \cite{Mitra:2023yjf}. 

This article's main objective is to examine the deflection angle and gravitational lensing by such WGEB wormholes. 
In order to accomplish this, we must solve geodesic equations for null trajectories that correspond to GEB and WGEB wormhole passages. The following are the geodesic equations in the GEB spacetimes: 
\begin{equation}
\frac{d^{2}t}{d\lambda^{2}} = 0 ,\label{eq:geodesic-1}
\end{equation}
\begin{equation}
\frac{d^{2}l}{d\lambda^{2}} - \f{l^{m-1}} {( b_{0}^{m} + l^{m})^{1 - 2/m}} ~ \le[ \le( \frac{d\theta}{d\lambda} \ri)^{2} + \sin^{2}\theta  ~\le( \frac{d\phi}{d\lambda} \ri)^{2} \ri] = 0 ,\label{eq:geodesic-2}
\end{equation}
\begin{equation}
\frac{d^{2}\theta}{d\lambda^{2}} + \frac{2~l^{m-1}}{\le( b_{0}^{m} + l^{m} \ri)} \frac{dl}{d\lambda} \frac{d\theta}{d\lambda} - \sin\theta \cos\theta \le( \frac{d\phi}{d\lambda} \ri)^{2} = 0 ,\label{eq:geodesic-3}
\end{equation}
\begin{equation}
\frac{d^{2}\phi}{d\lambda^{2}} + 2 \cot\theta \frac{d\theta}{d\lambda} \frac{d\phi}{d\lambda} + \frac{2l^{m-1}}{ \le( b_{0}^{m} + l^{m} \ri) } \frac{dl}{d\lambda} \frac{d\phi}{d\lambda} = 0 .\label{eq:geodesic-4}
\end{equation}

The geodesic equations for the WGEB model are as follows:
\begin{equation}
\frac{d^{2}t}{d\lambda^{2}} + 2 ~f'(y) ~\frac{dt}{d\lambda} ~\frac{dy}{d\lambda} = 0 ,\label{eq:geodesic-11}
\end{equation}
\begin{equation}
\frac{d^{2}l}{d\lambda^{2}} + 2~f'(y)~\frac{dl}{d\lambda}~\frac{dy}{d\lambda}  - \f{l^{m-1}}{\le( b_{0}^{m} + l^{m} \ri)^{1 - 2/m}} ~\le[ \le( \frac{d\theta}{d\lambda} \ri)^{2} + \sin^{2}\theta  ~\le( \frac{d\phi}{d\lambda} \ri)^{2} \ri] = 0 ,\label{eq:geodesic-22}
\end{equation}
\begin{equation}
\frac{d^{2}\theta}{d\lambda^{2}} + 2 ~f'(y) ~\frac{d\theta}{d\lambda} ~\frac{dy}{d\lambda} + ~\frac{2~l^{m-1}}{(b_{0}^{m} + l^{m})} ~\frac{d\theta}{d\lambda} ~\frac{dl}{d\lambda} - \sin\theta ~\cos\theta ~\le( \frac{d\phi}{d\lambda} \ri)^{2} = 0 ,\label{eq:geodesic-33} 
\end{equation}
\begin{equation}
\frac{d^{2}\phi}{d\lambda^{2}} + 2 ~f'(y) ~\frac{d\phi}{d\lambda} ~\frac{dy}{d\lambda} + 2 ~\cot\theta ~\frac{d\theta}{d\lambda} ~\frac{d\phi}{d\lambda} + \frac{2l^{m-1}}{ \le( b_{0}^{m} + l^{m} \ri) } ~\frac{dl}{d\lambda} ~\frac{d\phi}{d\lambda} = 0 ,\label{eq:geodesic-44}
\end{equation}
\begin{equation}
\frac{d^{2}y}{d\lambda^{2}} + f'(y)\, e^{2f(y)}\le[\le( \frac{dt}{d\lambda} \ri)^{2} - \le(\frac{dl}{d\lambda} \ri)^{2} - (b_{0}^{m} + l^{m})^{2/m} \le[ \le(\frac{d\theta}{d\lambda} \ri)^{2} + \sin^{2}\theta \le( \frac{d\phi}{d\lambda} \ri)^{2} \ri] \ri] = 0 .\label{eq:geodesic-55}
\end{equation}
In the above equations $\lambda$ is the affine parameter. The differences in geodesic equations of two cases (GEB and WGEB) are the additional geodesic equation for motion along the extra dimension $y$, as well as the presence of $y$ and $\dot{y}$ dependent terms in the right hand side of the Eqs. (\ref{eq:geodesic-11}) to (\ref{eq:geodesic-44}) . A full investigation of the time-like geodesics for the above scenarios has been reported in  \cite{Sharma:2022tiv}, which reveals the confinement of test particles along the extra dimension in the case of growing warp factor, and  runaway trajectories in presence of decaying warp factor. Here, we are interested in the deflection angle and gravitational lensing of the null trajectories.

\section{Null-Trajectories}
\label{sec:Null-Trajectories}
Since the geodesic equations are highly coupled nonlinear second-order differential equations, obtaining fully analytic solutions is generally difficult. 
Therefore, we use the constants of the motion that can be derived from the metric itself using corresponding Euler-Lagrange's equations and the geodesic constraint for the null geodesic, given by,
\begin{equation}
g_{\mu\nu}u^{\mu}u^{\nu} = 0 ,\label{eq:null_geodesic_constraint}
\end{equation}
where, $u^{\mu}$ is the velocity null-vector. 
Two scenarios might be considered interesting. One is the bending of light, relative to a distant observer, coming from the other side of the wormhole from our line of sight (to be discussed in Sec IV and V). 
Besides the standard lensing scenario involving a distant observer, one may also consider a local probe of the wormhole geometry, where light rays are emitted toward the throat and reflected back to the observer. Such returning trajectories may, in principle, encode information about the wormhole parameters through interference effects and winding numbers.
In the following, we discuss the null trajectories for both GEB and WGEB scenarios.

\subsection{Null rays in GEB Space-time}

According to Eq. (\ref{eq:null_geodesic_constraint}), the null geodesic constraint for GEB space-time is,
\begin{equation}
-\dot{t}^{2} + \dot{l}^{2}+ r^2 \left( \dot{\theta}^{2} + \sin^{2}\theta \, \dot{\phi}^{2}\right) = 0 .\label{eq:null_constraint_4D}
\end{equation} 
Where, an over-dot denotes derivative with respect to the affine parameter $\lambda$. From here on, we shall express $r(l)$ as $r$. The constants of motion corresponding to cyclic coordinates ($t, \phi$)  (when $\theta = \pi/2$), are respectively, 
\begin{equation}
\dot{t} = k ~~~\text{and}~~~ r^2 \dot{\phi} = h . \label{eq:constants_for_4D_GEB}
\end{equation}
Here, $k$ and $h$ are integration constants representing the conserved energy and angular momentum of light-like particles respectively. 
Thus the areal velocity ($r^2 \dot\phi$) is a conserved quantity as expected.
One can further write down the following relation, using Eqs. (\ref{eq:null_constraint_4D}) and (\ref{eq:constants_for_4D_GEB}), 
\begin{equation}
\dot{l}^{2}= k^{2} - \frac{h^{2}}{r^2} .\label{eq:null-trajectories}
\end{equation}
The equation of the photon trajectories (in the $l-\phi$ plane) can be written in terms of the so-called impact parameter, $b = h/k$, (which is defined as the perpendicular distance from the center of the wormhole {\em lens} to the projectile path) as follows,
\begin{equation}
\left( \frac{dl}{d\phi} \right)^{2} = r^2 \left( \frac{ r^2 }{b^{2}} - 1 \right) .\label{eq:null-trajectory2}
\end{equation}
Since, $r \ge b_0$, a photon coming from large (positive) $l$, will not reach the throat radius ($l=0$) if there exists a point $l_{0} > 0$ where $\frac{dl}{d\phi}$ or $\dot{l}$ vanishes. Then $l_{0}$ would be the point of closest approach (along the tortoise or the proper radial coordinate) given by the relation,
\begin{equation}
l_{0}^{} = \left( b^{m} - b_{0}^{m}\right)^{1/m} .\label{eq:l0_4D}
\end{equation}  
Further, using Eqs. (\ref{eq:null-trajectory2}), we arrive at,
\begin{equation}
 \frac{d\phi}{dl} = \pm \frac{1}{\sqrt{r^2 \left( \frac{r^2}{b^2} - 1\right) }} ,\label{eq:null-trajectory3}
 \end{equation}
 which is the key equation to compute deflection angle. Note that, for a point of return to exist, the impact parameter $b$ must be greater than $b_0$ (otherwise we have `passing' or `trapped' trajectories \cite{Sharma:2022tiv}). 
 Thus, choosing any particular $h/k$ is same as choosing the point of return (for a geometry with given $b_0$ and $m$). This also tells us that the radial geodesics with angular momentum $h<b_0k$ will always pass through the neck. 
 Note that, in principle one can determine the wormhole parameters from characteristic quasi-normal modes \cite{DuttaRoy:2019hij, Mitra:2023yjf}.
 We shall return to this equation with it's solution for the deviation angle, in Sec. \ref{sec:def}.
 Below we discuss the 5D scenario and effects of the warping factor.
\subsection{Null rays in WGEB Space-time}

The null trajectories in the WGEB space-time satisfies the null constraint,
  \begin{equation}
  e^{2f(y)} \left[ -\dot{t}^{2} + \dot{l}^{2}+ r^2 \left( \dot{\theta}^{2} + \sin^{2}\theta \dot{\phi}^{2}\right) \right] + \dot{y}^{2} = 0 .\label{eq:5D_null_constraint}
  \end{equation}
Thus the constants of motion for WGEB wormhole at $\theta = \pi/2$ are, say,
\begin{equation}
e^{2f(y)} \dot{t} = K ~~~\text{and}~~~ e^{2f(y)}r^2 \dot{\phi} = H .\label{eq:5d_constant}
\end{equation}
Note that the areal velocity ($r^2 \dot\phi$) is a conserved quantity in 5D scenario as well.
Matching the 5D and 4D scenarios at $y=0$, implies $H=h$ and $K=k$.
By using a similar approach as in the case of 4D model, we obtain the following equation corresponding to null geodesics (\ref{eq:geodesic-11}) to (\ref{eq:geodesic-55}),
\begin{equation}
\dot{l}^{2} = \frac{1}{e^{2f(y)}} \Big[ \frac{k^{2}}{e^{2f(y)}} - \frac{h^{2}}{e^{2f(y)} r^2} - \dot{y}^{2} \Big] .\label{eq:null-l-dot}
\end{equation}
This equation can be further simplified using  Eqs.~(\ref{eq:5D_null_constraint})  in  Eq.~(\ref{eq:geodesic-55}) as,
\begin{equation}
\frac{d^2 y}{d\lambda^2}+ f'(y)\left(\frac{dy}{d\lambda}\right)^2=0,
\label{eq:null_y}
\end{equation}
which after integration results in 
\begin{equation}
\dot{y}=C e^{-f(y)}, \la{eq:ydot}
\end{equation}
where $C$ is an integration constant (that quantifies momentum component, of a photon, along the warped extra dimension). One may solve Eq. \ref{eq:ydot}, for the growing and decaying warp factors, $f(y)=\pm \log \left[\cosh(\frac{y}{y_0})\right]$. This leads to
\be
y_+ = y_0 \sinh ^{-1}\big(\f{C}{y_0}\l + C_1\big)~~~~ \mbox{and}~~~~ y_- = y_0 \sinh^{-1} \tan \big(\f{C}{y_0}\l + C_1\big) .
\ee
These analytic results however are not of our current interest.
Moving ahead, Eq. \ref{eq:null-l-dot} further simplifies to
\begin{equation}
\dot{l}^{2} = \frac{1}{e^{4f(y)}} \Big[ k^2 - \frac{h^{2}}{r^2} - C^2 \Big] .\label{eq:null-l-dot-1}
\end{equation}
Therefore, in the 5D case, the point of return or the impact parameter along coordinates $r$ ($b_w$) and $l$ ($l_0$) are respectively given by,
\be
l_{0}=\left(b_w^{m}-b_0^m\right)^{1/m} ~~~~ \mbox{where} ~~~ b_{w} = \f{b}{\sq{1-\d^2}}~ , \label{eq:impact-5D}
\ee
where, we define the dimensionless parameter $\delta=C/k$, an initial-condition parameter of a null geodesic 
which parametrises the conserved extra-dimensional momentum and its effect on the projected null trajectory. Its observational imprint is encoded through the effective impact parameter $b_w$. 
Eq. (\ref{eq:impact-5D}) implies that the effective impact parameters vary with varying $\d$ for any given $h/k$. Further, for any given $\d$, $l_0=0$ corresponds to null rays satisfying $b/b_0 = \sq{1-\d^2}$ (thus $b$ can be less than $b_0$).

In a Kaluza–Klein (KK) interpretation \cite{Herrera-Aguilar:2014oua}, a non-zero $C$ may be viewed as an effective four-dimensional mass, but the present work adopts a purely geometric viewpoint and does not explicitly identify the projected null geodesics with massive photons.
In a thick-brane scenario, the electromagnetic field may be described by a localized four-dimensional zero mode, while additional bulk or KK modes can have nontrivial dependence on the extra-dimensional coordinate. The realization of $C\neq0$ as an electromagnetic signal therefore requires a specific localization/propagation mechanism for the electromagnetic sector. We consequently interpret the null rays with $\d\neq0$ here as five-dimensional null modes with nonzero extra-dimensional momentum, rather than assuming that ordinary brane-localized photons generically have $C\neq0$. 
Further, $C\neq0$, implies $\dot y\neq0$, and hence the corresponding five-dimensional null geodesic is not confined to the brane. This raises a distinction between the geometrical propagation of bulk null trajectories and the physical transmission of an electromagnetic signal between brane-localized source and observer. 
The deflection angle and related lensing quantities derived for ($C\neq0$) only characterize the geometry of the five-dimensional null geodesics; they do not, by themselves, establish that these trajectories correspond to observable electromagnetic signals between brane-localized source and observer. Such an interpretation requires an additional field-theoretic treatment of electromagnetic localization, propagation, and coupling to brane matter, and is left for future work.

We treat $\d$ as a free dimensionless parameter whose observational consequences are explored.
Note that, values of $\d$ approaching unity would correspond to geodesics carrying most of their conserved momentum into the bulk and are therefore expected to have limited observational relevance for brane-based observers. Accordingly, the regime $\d <<1$ is likely to be the most realistic from a phenomenological perspective.
However, deriving observational constraints on $\d$ would require a specific localisation mechanism (on thick branes) through a coupling of the electromagnetic field with background spacetime (that could lead to a localised zero mode and a continuum of KK tower) and a detailed effective four-dimensional field theory, which lie beyond the scope of the present work.

Let us explore if we can distinguish the effects of any wormhole parameter. Firstly, $b_0$ is a wormhole parameter independent of properties of the null trajectories. On the other hand, effect of warped extra dimension, $0 \leq \d^2 \leq 1$, can vary.
For $\d^2 << 1$, we can further approximate the above equations as,
\begin{equation}
b_{w} \sim b \le(1+ \f{\d^2}{2}\ri) ~~~~ \mbox{or}~~~~  l_{0} \sim \left[b^m \le(1+ \f{m}{2}\delta^2 \ri) -b_0^m\right]^{1/m}. \la{eq:l05-2}
\end{equation}
Thus, presence of $\d$ leads to a larger impact parameter for any given $h/k$ and variation in $\d$ results in different closest approach of light rays that have same $h/k$. Similarly, different combinations of $h/k$ and $\d$  may lead to the same impact parameter $b_w$. A relative uncertainty or error in the impact parameter thus induced by the extra dimension is given by. 
\be
\f{\triangle b_{w}}{b_w} \sim  \d\D  \d . \la{eq:error_r0}
\ee
Eq. (\ref{eq:error_r0}) shows that the effect of the extra dimension manifests through a $\d$-dependent change in the effective impact parameter, which can lead to different angular radii for rays with different extra-dimensional momenta (discussed later). 
In the following, we analyse the null trajectories further in order to distinguish the effect of the wormhole parameters.

\subsection{The photon sphere}

The equation of null trajectories in warped background on the $l - \phi$ plane, in 5D are given by
\be
 \frac{d\phi}{dl} = \pm \frac{1}{\sqrt{r^2 \left( \frac{r^2}{b_w^2} - 1\right) }}, \label{eq:null-trajectory-5D}
\ee
which looks functionally similar to Eq. (\ref{eq:null-trajectory3}) corresponding to the 4D scenario. 
Since every lensing observable is ultimately derived from the null geodesic or above equation for deflection angle, the complete five-dimensional lensing formalism is obtained exactly from the four-dimensional one through the mapping $b \ra b_w = b/\sq{1-\d^2}$. 
Hence, every observable $\mc{O}$ computed in the paper satisfies the following mapping between 4D and 5D quantities--
\be
\mc{O}_{5D}(b,\d) =  \mc{O}_{4D}(b_w).   \la{eq:map}
\ee
Thus, dependence of $b_{w}$ on $\d$ leads to parameter degeneracy among trajectories and observables.
Eq. (\ref{eq:null-trajectory-5D}) suggests, in general, there are three types of null trajectories-- Trapped, Returning and Crossing trajectories \cite{Sharma:2022tiv}. 
Depending on whether the point of return ($l_0$) is zero, positive or imaginary, one can characterise these trajectories as: (a) trapped at the throat, (b) ones that return before reaching the throat and (c) the ones that cross the throat and reaches the other side.
Thus, we get the following classifications.
For null trajectories in the 5D-WGEB wormhole scenario-- (a) $h = b_{0}  \sqrt{k^{2} - C^2} $ $\Longrightarrow$ trapped, (b) $h > b_{0}  \sqrt{k^{2} - C^2}$ $\Longrightarrow$ returning, and (c) $h < b_{0}  \sqrt{k^{2} - C^2}$ $\Longrightarrow$ crossing trajectories. Putting $C=0$, one recovers four-dimensional version of these classification.

{\em Trapped} trajectories essentially lead to the formation of a photon sphere at the throat radius.
 Photon sphere is a region where a photon can travel in a circular orbit i.e. it is the region where both $\dot{l}$ and $\ddot{l}$ are zero. Using Eq. (\ref{eq:null-trajectories}) and geodesic equations, we get the following conditions for location of photon sphere in GEB space-time:
\be
l = \left(b^{m} - b_{0}^{m} \right)^{1/m} = 0 . \la{eq:photon-sphere-condition1} 
\ee

For WGEB wormhole, the conditions for photon sphere, which is again located at $l=0$ (using Eq. (\ref{eq:null-l-dot}) or (\ref{eq:null-l-dot-1}), implies
\be
b= b_0 \sq{1-\d^2}. \la{eq:5d-ps-condition}
\ee

Since $\d$ can vary, given any fixed $b_0$, Eq. (\ref{eq:5d-ps-condition}) is satisfied by different  combinations of $h/k$ and $\d$. 
Hence, the parameter $\d$ does not alter the coordinate radius of the throat or the geometric location of the photon sphere. Instead, for a fixed projected impact parameter ($b=h/k$), different values of $\d$ lead to different effective impact parameters ($b_w$) and consequently to different closest-approach distances ($l_0$). 
The resulting variation should therefore be interpreted as a family of trajectories or image positions corresponding to different values of the conserved extra-dimensional momentum.
If there exists another phenomenon that connects $h/k$, $b_0$ and $\d$, in principle one can constrain $C$ by eliminating $b_0$. 
Presence of photon sphere in a given wormhole passage, would give rise to {\em relativistic images} near throat radius \cite{Virbhadra:1999nm}. 
In the next section, we analyse deflection of  null trajectories and gravitational lensing in detail.

\section{Deflection-Angle}\la{sec:def}

We now compute the total deflection angle of null rays that approach the wormhole from asymptotic infinity, reach a point of closest approach $l_0$, and return to infinity.
Let us consider that, on the equatorial plane ($\theta = \pi/2 $), incident direction of photon is $\phi_{\infty}$ (at $l = \infty $) and the direction of photon at closest approach (at $l_{0}$) is $\phi_{0}$. Then total change in $\phi$ can be given by, 
\begin{equation}
\Delta \phi = 2 \Big| \phi_{0} - \phi_{\infty} \Big| . 
\end{equation}
The pre-factor $2$ appears because we are first considering the change in $\phi$ for decreasing $l \in (\infty, l_{0})$ and then for increasing $l \in (l_{0}, \infty)$, second one is just time reversal of the first one. If the  trajectory of the photon were a straight line, then deflection of the photon orbit from the straight line is (for a complete derivation of the deflection angle, see \cite{Weinberg:1972kfs}),
\begin{equation}
\alpha(l_{0}) = 2|\phi_{0} - \phi_{\infty} | - \pi .\label{eq:deflection1}
\end{equation} 
Below we compute this deflection angle for both 4D and 5D scenarios.
\subsection{The 4D-GEB scenario}

Substituting the trajectory equation into the standard definition of the deflection angle yields the expression of deflection angle for 4D-GEB (without extra dimension) space-time as,
\begin{equation}
{}^{4}\alpha\left( l_{0} \right) = - \pi + 2 \int^{\infty}_{l_{0}} \frac{dl}{\sqrt{\left( b_{0}^{m} + l^{m} \right)^{2/m} \left[ \frac{\left( b_{0}^{m} + l^{m} \right)^{2/m}}{\left( b_{0}^{m} + l_{0}^{m} \right)^{2/m}} - 1 \right]}} .\label{eq:4d-da}
\end{equation}
Few general conclusions can be drawn on physical grounds about what to expect from Eq. (\ref{eq:4d-da}). There are two length scales involved -- $b_0$ and $l_0$. Their interplay further depends on the (dimensionless) steepness parameter $m$. The deviation angle is expected to be larger if $l_0 \sim b_0$. In increasing $m$-geometries, as the neck flattens out quickly for $r > b_0$, trajectories with $l_0 > b_0$  will have smaller deviations. From measurement of deviation angle alone, a far away observer can not determine both $b_0$ and $l_0$. 
Below we write the integral in Eq. (\ref{eq:4d-da}) in a more convenient form in terms of $x=l/b_0$ and $x_0=l_0/b_0$ as, ${}^{4D}\alpha\left( l_{0} \right) = - \pi + 2I$, where
\begin{equation}
I = \int^{\infty}_{x_{0}} \frac{dx}{\sqrt{\left( 1 + x^{m} \right)^{2/m} \left( \frac{\left( 1 + x^{m} \right)^{2/m}}{\left( 1 + x_{0}^{m} \right)^{2/m}} - 1 \right)}} .\label{eq:4d-da-x}
\end{equation}
The integrand diverges as $x\ra x_0$.
For $m = 2$ geometry (4D-EB), Eq. (\ref{eq:4d-da}) or (\ref{eq:4d-da-x}) can be integrated analytically in terms of complete elliptic integral of the first kind, and leads to \cite{Bhattacharya:2010zzb, Nakajima:2012pu, Tsukamoto:2012xs},  
\begin{equation}
{}^{EB}\alpha(l_{0}) = -\pi + 2 \sqrt{1 + \f{1}{x_{0}^2}} ~\mbox{Elliptic}K \Big(- \frac{1}{x_{0}^{2}} \Big). \label{eq:deflection-angle-EB1}
\end{equation}
The right hand side of Eq. (\ref{eq:deflection-angle-EB1}) diverges in the limit $x_0\ra 0$.
The divergence of the deflection angle as $l \ra 0$ indicates the onset of strong gravitational lensing. In this regime, photons execute multiple revolutions around the throat before escaping to infinity, leading to the formation of relativistic images and large winding numbers.

Similarly, changing to circumferential radial coordinate $r$, integral (\ref{eq:4d-da-x})  becomes
\begin{equation}
I = \int^{\infty}_{b}dr \frac{b~ r^{m-2}}{(r^m - b_0^m)^{\f{m-1}{m}} \sq{r^2-b^2} } .\label{eq:4d-da-r}
\end{equation}
One may also define $u=1/r$ \cite{Tsukamoto:2012xs}, that modifies the Eq. \ref{eq:4d-da-r} into
\begin{equation}
I = \int^{b^{-1}}_0 \frac{b~du}{( 1 - b_0^m u^m)^{\f{m-1}{m}} \sqrt{ 1- b^2 u^2 }} .\label{eq:4d-da-u}
\end{equation}
These integrals can be solved exactly \cite{Tsukamoto:2012xs, Bhattacharya:2010zzb, Nakajima:2012pu} for $m=2$ and we get
\begin{equation}
{}^{EB}\alpha(b) = -\pi + 2 ~\mbox{Elliptic}K \Big(\frac{b_0^2}{b^2} \Big). \label{eq:deflection-angle-EB2}
\end{equation}
$K(y)$ diverges in the limit $y\ra 1$, which implies strong lensing.
For $m > 2$ we have numerically solved for the deflection angle and plotted the results later.
We derive approximate results in weak and strong lensing limits for $m \geq 2 $ is presented below.

\subsubsection{Weak lensing}
The limit $b_0/l_0 \ra 0$ ($b_0/b \ra 0$) or weak lensing, leads to a well known result for EB model \cite{Bhattacharya:2010zzb, Nakajima:2012pu, Tsukamoto:2012zz, Tsukamoto:2012xs},
\begin{equation}
{}^{EB}\alpha_{weak} =  \frac{\pi b_{0}^{2}}{4 l_0^{2}}  - \frac{7\pi b_{0}^{4}}{64 l_0^{4}} + O \le( \frac{b_{0}^{6}}{ l_0^{6}}\ri) =  \frac{\pi b_{0}^{2}}{4 b^{2}}  + \frac{9\pi b_{0}^{4}}{64 b^{4}} + O \le( \frac{b_{0}^{6}}{ b^{6}}\ri) . \label{eq:def-EB2-limit}
\end{equation}
The deviation angle tends to infinity as $l_0 \ra 0$ (or $b \ra b_0$ i.e. strong lensing), which implies that returning trajectories that goes too close to the neck does so by whirling around the neck many times and then return to $l \ra \infty$ there after. 
Note that for some $l_0$, the right hand side of Eq. \ref{eq:deflection-angle-EB1} may become multiple of $2\pi$, thus providing means to measure the `winding number' (perhaps) using interference phenomenon. 
Interestingly, for larger $m$-geometries, approximate results can be derived from Eq. (\ref{eq:4d-da-x}) that would support the numerical results to follow. 
In the limit $x_0>>1$ or $b>>b_0$ (weak lensing), Integral (\ref{eq:4d-da-x}) can be written as
\be
I =  \le[\sum_{p=0}^{\infty}  \binom{\f{1}{m}}{p} \f{1}{x_0^{pm}}\ri] \int^{\infty}_{x_{0}}dx \le( \f{x_0/x}{\sq{x^2-x_0^2}} \ri)
\le[\sum_{q=0}^{\infty}  \binom{\f{-1}{m}}{q} \f{1}{x^{q m}} \ri] 
\le[1 + \sum_{k=1}^{\infty}  \binom{\f{2}{m}}{k} \f{ \f{x^2}{x^{km}} -  \f{x_0^2}{x_0^{km}}}{(x^2-x_0^2)} \ri]^{-\f{1}{2}}  ~~  \la{eq:master-int}
\ee
This integral can be exactly evaluated at each order $x_0^{-km}$ (after one more binomial expansion of the squared root term at the end). 
Thus up to order $x_0^{-2m}$ (for a detailed derivation, see Appendix \ref{app:WL-GEB}), we get
\ba
{}^{GEB}\alpha_{weak} &\simeq & \a_m \f{b_0^{m}}{l_0^m } + \a_{2m} \f{b_0^{2m}}{l_0^{2m} } + O \le(\f{b_0^{3m}}{l_0^{3m} }\ri) ,\la{eq:def-GEB-weak} \\ 
\mbox{where}~~~~\a_m &=& \f{2 }{ m} \le[\f{\pi}{4} + \le(\f{\sq{\pi}}{2} - \f{\sq{\pi}}{m}\ri) \f{\G(\f{m+1}{2})}{\G(\f{m}{2})} \ri] ,  \label{eq:alpha-m-weak}\\
\mbox{and}~~~~\a_{2m} &=& -\f{3\pi}{2m^2} + \f{4\sq{\pi}}{m^2} \f{\G(\f{m+1}{2})}{\G(\f{m}{2})} - \f{\sq{\pi}}{m^3} \f{\G(m+\f{3}{2})}{\G(m-1)}  . \label{eq:alpha-2m-weak}
\ea
Eq. \ref{eq:def-GEB-weak} reproduces Eq. (\ref{eq:def-EB2-limit}) for $m=2$ and predicts the deflection angles for higher-$m$ geometries.
Thus Eq. (\ref{eq:def-GEB-weak}) is one of the key results (derived from Eq. (\ref{eq:master-int})) we report in this article.
This shows that for GEB wormholes, $\a_m$ decreases monotonically ($\a_2\simeq 0.785, \a_4 \simeq0.69, \a_6\simeq0.59, \a_8\simeq0.52, \a_{10}\simeq0.47$) with increasing `$m$' (due to decrease in curvature at a particular $x > x_0$ with increasing $m$). 

\subsubsection{Strong lensing}

\begin{figure}[!htbp]
\centering
\subfigure[$m=4$]{\includegraphics[width=2.75in,angle=360]{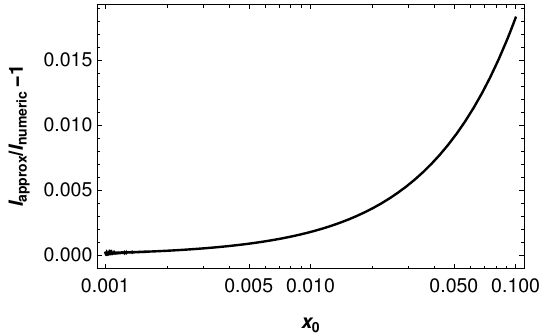}\label{subfig:Inumeric1}}
\subfigure[$m=6$]{\includegraphics[width=2.75in,angle=360]{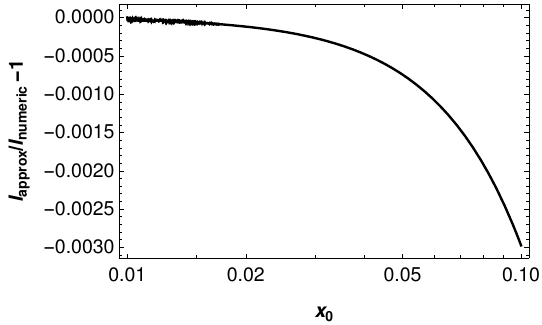}\label{subfig:Inumeric2}}
\caption{\raggedright Plot of $(\f{I_{approx}}{I_{numeric}}-1)$ as a function of $x_0$ , where $I_{approx}$ is the asymptotic expression ${}^{GEB}I_{x_0 \ra 0}$ given in Eq.~(\ref{eq:SL-approx-GEB-I}) and $I_{numeric}$ is numerical integration given in Eq.~(\ref{eq:4d-da-x}). The plots show that in the limit $x_0 \rightarrow{0+}$, the integral given in Eq.~(\ref{eq:4d-da-x}) approaches the asymptotic expression given in Eq.~(\ref{eq:SL-approx-GEB-I}). }
\label{fig:int-approx}
\end{figure}

Let us analyse strong lensing where the trajectories are flying {\em close} to the neck ($x_0 << 1$). 
In the limit $x_0=l_0/b_0 \ra 0$ ($r \ra b_0$), Eq. \ref{eq:deflection-angle-EB1} gives, to dominant order,
\be
{}^{EB}\a_{x_0 \ra 0} = -\pi + \log \le(\f{16}{x_0^2}\ri) + \f{x_0^2}{4} \le[2 + \log \le(\f{16}{x_0^2}\ri) \ri] + \mbox{H.O.T} \approx \log \le(\f{b_0^2}{l_0^2}\ri) . 
\la{eq:SL-limit-EB}
\ee
In terms of the radial coordinate or the impact parameter $b$, the above expression can be approximated as (keeping only the divergent term)
\be
{}^{EB}\a_{b \ra b_0} \approx  - \log\le(\f{b^2}{b_0^2} -1 \ri)  . \la{eq:SL-limit-EB-r}
\ee
Thus there is a logarithmic divergence for the total deflection angle for strong lensing.
Note that for any given $m$ and $x_0<<1$, in Eq. (\ref{eq:4d-da-x}) the divergent contribution to the integral is dominated by the values  near $x\sim x_{0+}$. Eq. (\ref{eq:4d-da-x}) can be simplified to (for detailed derivation see Appendix \ref{app:SL-GEB}),
\be
I_{x_0 \ra 0} \sim  \sq{\f{m}{2}}\int^{1}_{x_{0}} dx \f{1}{\sq{x^m - x_0^m}} + \mbox{Finite terms}. \la{eq:SL-approx} 
\ee
Thus for $m=2$, keeping only the divergent term, we get,
\be
{}^{EB}I_{x_0 \ra 0} = \lim_{x_0\ra 0+} \tanh^{-1} \sq{1-x_0^2} \approx -\log (x_0)  , \la{eq:SL-approx-EB-I}
\ee
which leads to strong lensing deflection angle ($\a = -\pi + 2I$) given by
\be
{}^{EB}\a_{SL} \approx -2\log (x_0) = -\log \le(\f{b^2}{b_0^2}-1\ri) , \la{eq:SL-approx-EB}
\ee
which matches the leading and diverging $x_0$-dependent term in Eq. (\ref{eq:SL-limit-EB}). For $m>2$, we arrive at
\be
{}^{GEB}I_{x_0 \ra 0} \simeq  \sq{\f{1}{2m}} B\le(\f{1}{2}-\f{1}{m}, \f{1}{2}\ri)x_0^{1-\f{m}{2}}. \la{eq:SL-approx-GEB-I}
\ee 
Where $B(p,q)$ is the beta function.
To explicitly verify the limiting behaviour $ {}^{GEB}I_{x_0 \ra 0}$ as given in Eq.~(\ref{eq:SL-approx-GEB-I}), we numerically evaluate the original deflection-angle integral (Eq.~\ref{eq:4d-da-x}) for representative higher-order geometries $m=4,6$ and compare it with the above asymptotic expression for progressively smaller $x_0$ until the numerical integration becomes unstable. We demonstrate this comparison graphically in  Fig.~\ref{fig:int-approx}. We find that the ratio of the numerical result to the leading asymptotic term approaches unity as $x_0 \to 0^+$, confirming the predicted power-law scaling. Opposite signs of ($I_{approx}/I_{numeric}-1$) for $m=4$ and $m=6$ case can also be understood from the asymptotic expansion (see Appendix \ref{app:SL-GEB}).
Thus, for the deviation angle in the strong lensing limit in ($m>2$) geometries, we have (retaining only the leading term)
\be
{}^{GEB}\a_{SL} \sim \sq{\f{2}{m}} B\le(\f{1}{2}-\f{1}{m}, \f{1}{2}\ri)x_0^{1-\f{m}{2}} = \sq{\f{2}{m}} B\le(\f{1}{2}-\f{1}{m}, \f{1}{2}\ri)\le(\f{b^m}{b_0^m}-1\ri)^{\f{1}{m} - \f{1}{2}}. \la{eq:SL-approx-GEB}
\ee
Eq. (\ref{eq:SL-approx-GEB-I}) or Eq. (\ref{eq:SL-approx-GEB}) provides another main result of this article.
It is important to note that Eq. (\ref{eq:SL-approx-EB-I}) cannot be recovered from Eq. (\ref{eq:SL-approx-GEB-I}) by taking the $m \ra 2^+$ limit as the full asymptotic expansion is non-uniform in the joint limit $m \ra 2^+, x_0 \ra 0^+$ (see Appendix \ref{app:limit-m-2} for details).

The analytic approximations suggest that, for any given $x_0<<1$, we expect larger deflection (and winding number) with increasing $m$ (because of the power law term in the denominator) which supports the numerical results presented in Fig.~\ref{fig:deflection angle} below. 
This behavior can be physically understood as follows.
All derivatives of $r(l)$ up to the order $(m-1)$ vanish at the throat. Thus in higher $m$-geometries, the function $r(l)$ has a higher-order stationary point at $l=0$. 
The throat becomes progressively `flatter' in terms of derivatives, even though it is geometrically steeper away from the immediate vicinity (because the first nonzero term is $r(l) - b_0 \sim l^m$).
A local expansion of the effective photon potential ($V(r)\sim r^{-2}$) near the throat (a maxima) would be an expansion in derivatives of $r(l)$.
Thus, the effective potential develops a higher-order stationary maximum at the throat for larger $m$, leading to the altered strong-deflection divergence.
For fixed closest-approach parameter $x_0$, these implies a corresponding increase in the possible winding number (however, it is the $\a$ modulo $2\pi$ that counts).
We thus have two equations-- (\ref{eq:def-GEB-weak}) and (\ref{eq:SL-approx-EB} / \ref{eq:SL-approx-GEB}) (modulo $2\pi$) among the parameters $l_0/b_0$ or $b/b_0$, $m$ and deflection angles for weak and strong lensing. 
Thus, if the projected impact parameter $h/k$ can be independently constrained and deflection measurements are available over a suitable range of impact parameters, the functional dependence and amplitude can in principle constrain $m$ and $b_0$.

\subsection{5D-WGEB scenario}

In the 5D-WGEB space-time, one can write the expression of deflection angle  using Eqs. (\ref{eq:null-trajectory-5D}) and (\ref{eq:deflection1}) as
\ba
{}^{5}\alpha\left( l_{0} \right) &=& - \pi + 2 \int^{\infty}_{l_{0}} \frac{dl}{\sqrt{\left( b_{0}^{m} + l^{m} \right)^{2/m} \left[ \frac{\left( b_{0}^{m} + l^{m} \right)^{2/m}}{\left( b_{0}^{m} + l^m_{0} \right)^{2/m}} - 1 \right]}}. \label{eq:5d-da}
\ea
where, $l_0$ is the point of return along coordinate-$l$ in 5D scenario as given by Eq. (\ref{eq:impact-5D}). 
One could clearly see from Eq. (\ref{eq:4d-da}) and Eq. (\ref{eq:5d-da}) that the functional dependence of $\a$ on $l_0/b_0$ and $m$ are same in both 4D and 5D scenarios. 
However, the counter part of Eq. (\ref{eq:4d-da-r}), in 5D, is given by
\begin{equation}
I = \int^{\infty}_{b_w}dr \frac{b_w~ r^{m-2}}{(r^m - b_0^m)^{\f{m-1}{m}} \sq{r^2-b_w^2} } .\label{eq:5d-da-r}
\end{equation}
where $b_w$ is given by Eq. (\ref{eq:impact-5D}).
In the limit, $b_w \ra b_0$, the integral $I$ above diverges in the same way as in the 4D case.
Interestingly, for any given $b=h/k$, due to the presence of $\d$ the impact parameter ($b_w$) is greater than $h/k$ i.e. $b_w > b$ and thus the above integral results in a smaller amount of total deflection angle.

All the analytic results of previous subsection remain functionally similar, with only one modification, i.e. replacing the impact parameter $b=h/k$ with $b_w$ as given by Eq. (\ref{eq:impact-5D}).
Thus solution for $m=2$ case will be given by Eq. 
\begin{equation}
{}^{EB}\alpha(b_0/b_w) = -\pi + 2 ~\mbox{Elliptic}K \Big(\frac{b_0^2}{b_w^2} \Big). \label{eq:deflection-angle-WEB2}
\end{equation}
For weak lensing ($b_w >>b_0$) Eq. (\ref{eq:deflection-angle-WEB2}) gives
\begin{equation}
{}^{WEB}\alpha_{WL} \sim  \frac{\pi b_{0}^{2}}{4 b_w^{2}} + O\le( \frac{b_{0}^{4}}{ b_w^{4}}\ri) \sim \frac{\pi b_{0}^{2}}{4 b^{2}}(1-\d^2). \label{eq:def-WEB2-wl}
\end{equation}
For general $m$, we have, using Eq. (\ref{eq:def-GEB-weak}),
\begin{equation}
{}^{WGEB}\alpha_{WL} \sim   \f{2 }{ m} \le[\f{\pi}{4} + \le(\f{\sq{\pi}}{2} - \f{\sq{\pi}}{m}\ri) \f{\G(\f{m+1}{2})}{\G(\f{m}{2})} \ri]\f{b_0^{m}(1-\d^2)^{m/2}}{b^m }. \label{eq:WGEB-wl}
\end{equation}
Thus presence of the extra  dimension decreases the lensing angle and different combination of $b$ and $\d$ would essentially lead to same $\a$.
Similarly, for strong lensing ($b_w \ra b_0$), using Eq. (\ref{eq:impact-5D}) in Eq. (\ref{eq:SL-limit-EB}), for $m=2$ we have (keeping only the divergent term)
\be
{}^{WEB}\a_{SL} \approx - \log\le[ \f{b^2}{b_0^2(1-\d^2)} -1 \ri] . \la{eq:WEB-sl}
\ee
And for $m>2$ we get (upto the leading divergent term),
\be
{}^{WGEB}\a_{SL} \approx  \sq{\f{2}{m}} B\le(\f{1}{2}-\f{1}{m}, \f{1}{2}\ri)\le(\f{b^m}{b_0^m(1-\d^2)^{m/2}}-1\ri)^{\f{1}{m}-\f{1}{2}} . \la{eq:WGEB-sl}
\ee
Though we have shown explicitly how $\d$ appears in the deflection angle, one essentially measures the impact parameter $b_w = b/\sq{1-\d^2}$. 
As mentioned above, for any given $b=h/k$, the logarithmic term decreases with increasing $\d$.
Although the role of $\d$ is clearly different in Eqs. (\ref{eq:def-WEB2-wl}) and (\ref{eq:WEB-sl}), it is difficult to take advantage of this difference to distinguish the effect of the extra dimension unless one can measure $b=h/k$.
Eq. (\ref{eq:impact-5D}) or Eq. (\ref{eq:l05-2}) suggest that for known values of $h/k$, $b_w$ and $m$, one can constrain $\d$.

\begin{figure}[!htbp]
\centering
\subfigure[$m=2, \delta= 0.1$]{\includegraphics[width=1.9in,angle=360]{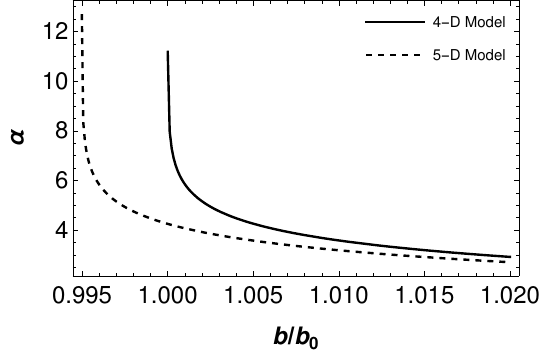}\label{subfig:p1}}
\subfigure[$m=2, \delta= 0.05$]{\includegraphics[width=1.9in,angle=360]{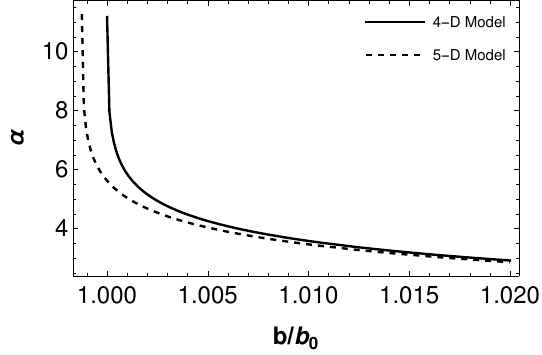}\label{subfig:p05}}
\subfigure[$m=2, \delta= 0.03$]{\includegraphics[width=1.9in,angle=360]{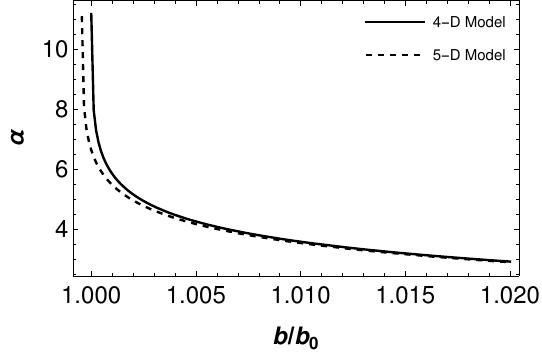}\label{subfig:p03}}
\subfigure[$m=4, \delta= 0.1$]{\includegraphics[width=1.9in,angle=360]{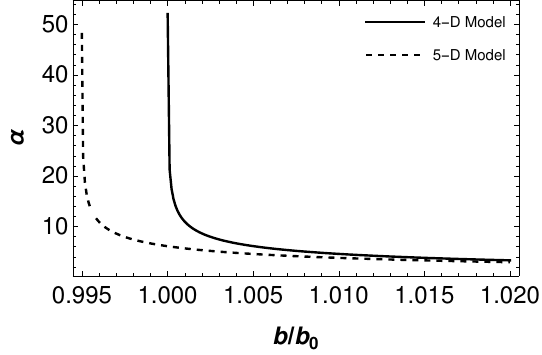}\label{subfig:p1m4}}
\subfigure[$m=4, \delta= 0.05$]{\includegraphics[width=1.9in,angle=360]{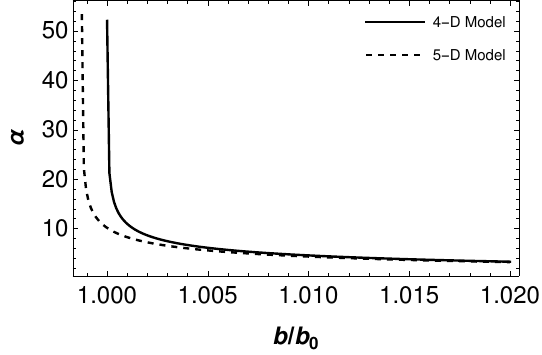}\label{subfig:p05m4}}
\subfigure[$m=4, \delta= 0.03$]{\includegraphics[width=1.9in,angle=360]{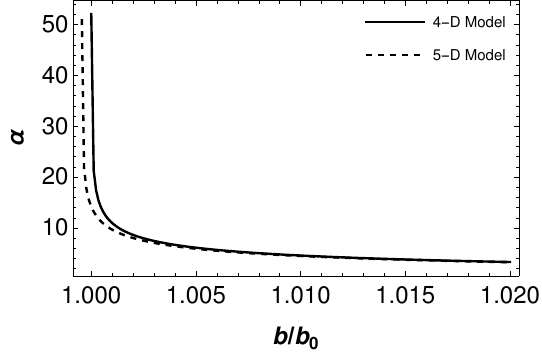}\label{subfig:p03m4}}
\subfigure[$m=6, \delta= 0.1$]{\includegraphics[width=1.9in,angle=360]{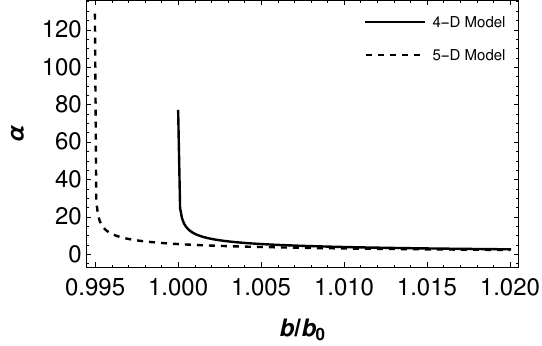}\label{subfig:p1m6}}
\subfigure[$m=6, \delta= 0.05$]{\includegraphics[width=1.9in,angle=360]{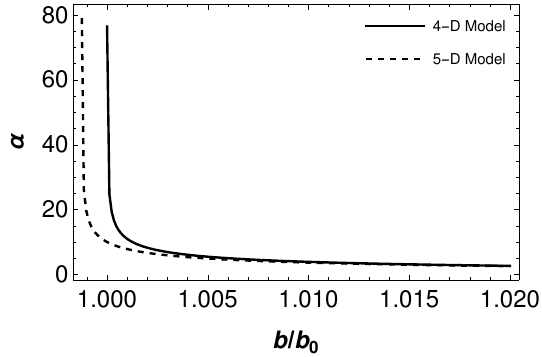}\label{subfig:p05m6}}
\subfigure[$m=6, \delta= 0.03$]{\includegraphics[width=1.9in,angle=360]{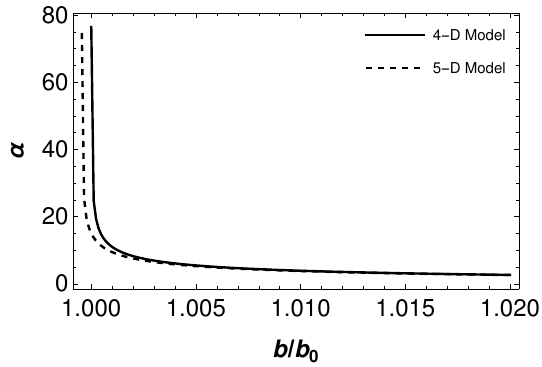}\label{subfig:p03m6}}
\caption{\raggedright Plot of deflection angles $\alpha$ as a function of the normalized impact parameter $b/b_0$ of the 4D GEB and 5D WGEB wormholes. The left-to-right panels show decreasing values of $\d$, while the top-to-bottom panels correspond to increasing values of the steepness parameter $m$. The figures demonstrate that as $\d$ decreases the minimum value of $b=h/k$ ($b_{min}$), for which the deflection angle diverges, shifts towards right in the figures, meaning $b_{min}/b_0 \ra 1$. On the other hand, as $m$ increases the divergence of deflection angle near $b_{min}/b_0$ increases although the value of $b_{min}/b_0$ remains almost fixed. However $h/k$ is a {\em non-observable} in lensing data.
}
\label{fig:deflection angle}
\end{figure}

In Figure~\ref{fig:deflection angle}, we plot the deflection angle $\alpha$ as a function of $b/b_0$ (for both the 4D and the 5D scenario) with different values of $m$ and $\delta$, where $b=h/k$, the ratio of energy and angular momentum of the null rays. In 5-D case, the impact parameter is $b_w= b/\sqrt{1-\delta^2}$.
We note that in the 5D scenario the minimum value of $b$, for which the deflection angle diverges, can be less than $b_0$, the throat radius in the 4-D scenario. 
This reduction of the critical value of $b=h/k$ below $b_0$ should not be interpreted as a reduction of the throat or photon-sphere radius; the latter remains $r_{\rm ph}=b_0$.
As we increase $\delta$ the effect is more prominent. This effect is independent of the value of $m$. However, as we increase $m$ the divergence of deflection angle becomes steeper.  
The numerical results are consistent with the analytic arguments presented earlier.
\begin{figure}[!htbp]
\centering
\includegraphics[scale=0.8]{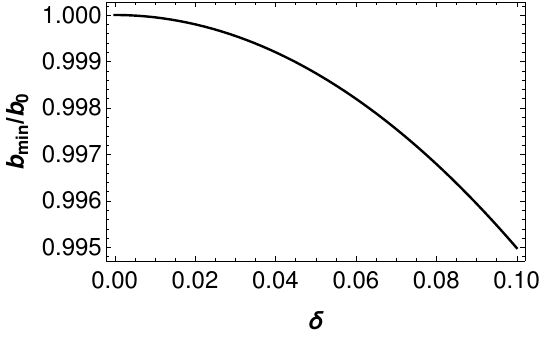}
\caption{The ratio $h/k=b_{min}$ of photon's angular momentum to it's energy (scaled with respect to $b_0$), corresponding to fixed closest approach ($l_0$) is plotted as function of $\delta$.  However $h/k$ is a {\em non-observable} in lensing data.}
\la{fig:b_min}
\end{figure}
In Fig.~\ref{fig:b_min}, we plot the variation of the minimum value of $b$ as the function of $\delta$ which is consistent with Figure~\ref{fig:deflection angle}. 
For an astrophysical observer, perhaps a potentially useful probe of these effects is gravitational lensing, which we analyse in the next section. 

\section{Gravitational Lensing}

\label{sec:lensing}
\begin{figure}[!htbp]
\centering
\includegraphics[scale=0.3]{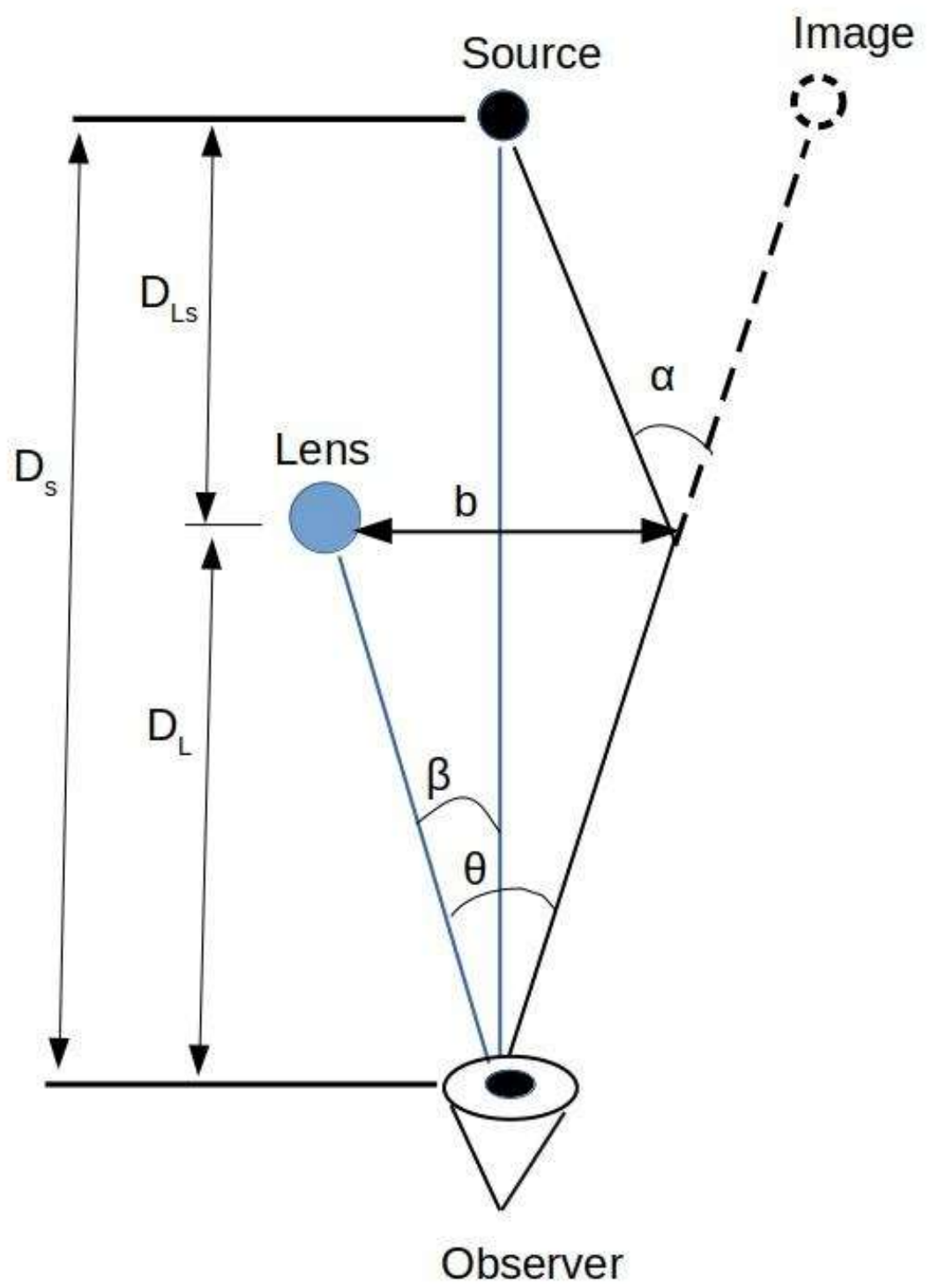}
\caption{Schematic diagram of gravitational lensing by a point source.}
\la{fig:lensing}
\end{figure}

The relation between Einstein-ring and relativistic-ring observables can provide a model-discriminating diagnostic for wormhole and black-hole lens models, as argued in \cite{Tsukamoto:2012xs}. 
For simplicity, we first analyse the light rays that are deflected by an angle less than $2\pi$ as these are the rays that form the so-called primary image. 
Figure \ref{fig:lensing} illustrates the standard lensing configuration involving the source (e.g. a star), lens (a wormhole or any compact massive object) and observer. The distances $D_{L}$, $D_{S}$ and $D_{LS}$ are distances between `observer-lens', `observer-source', and `lens-source', respectively. 
Here, $\theta$ and $\beta$ are the angles between `lens-image' and `lens-source'. 
Angle of deflection is represented by $\alpha$, and  $b$ is the impact parameter (or $l_0$ that depends on $\d$). 
Now, using basic geometrical law,
one can write the following set of identities \cite{Hartle:2003yu},
\begin{equation}
\theta = \frac{b}{D_{L}}, \alpha = \frac{D_{SI}}{D_{LS}}, \psi = \frac{D_{SI}}{D_{S}} . \label{eq:theta}
\end{equation}
Here, $\psi = (\th - \b)$ is the angle between source ($S$), observer and image ($I$), and $D_{SI}$ is the arc-length between the same. 
Thus, one obtains the general lensing equation, using Eqs. (\ref{eq:theta}) and Fig. \ref{fig:lensing}, as \cite{Virbhadra:1999nm}
\begin{equation}
\beta = \frac{b}{D_{L}} - \frac{D_{LS}}{D_{S}} \alpha ~~ ~~\text{for} \left( l_{0}>0 \right). \label{eq:gen-lensing-eq}
\end{equation}
We can now use the expressions of $\a$ derived in the previous section analyse different cases of weak and strong field limit. For the 5D scenarios one must replace $b$ by the effective impact parameter $b_w$  and corresponding deflection angle.
In 4D, in the weak field limit ($b >> b_0$), impact parameter, $b$ would be approximately same as $l_{0}$. Thus  the general lensing equation approximates to
\begin{equation}
\beta = \frac{l_{0}}{D_{L}} - \frac{D_{LS}}{D_{S}} \alpha_{WL}, ~~\text{for} ~~ l_{0} = (b^m - b_0^m)^{\f{1}{m}} >0  .\label{eq:lensing-eq-1}
\end{equation}
Note that images will be formed on the other side of the wormhole (for $l_{0} < 0$) following a similar manner. However, the direction of deflection angle  would be of opposite sign as given by
\begin{equation}
\beta = \frac{l_{0}}{D_{L}} + \frac{D_{LS}}{D_{S}} \alpha_{WL}, ~~\text{for} ~~ l_{0} = -(b^m - b_0^m)^{\f{1}{m}}<0 .\label{eq:lensing-eq-2}
\end{equation}
Interestingly, in the Schwarzschild lensing \cite{1992grlensing}, the deflection angle is an odd function of the impact parameter and, therefore, one of the above two equations is sufficient to explain both the situations ($l_{0} > 0$ and $l_{0} < 0$). 
More on the standard aspects of lensing equations are given in Appendix \ref{app:ang}.
In the following, we analyse the distinguishing features of Einstein radius, Einstein ring and so-called winding numbers specific to GEB and WGEB scenarios.

\subsection{Angular radii for non-relativistic and relativistic Einstein rings }

Perfect alignment of the source, lens, and observer ($\b=0$) leads to the formation of an Einstein ring (or Einstein relativistic rings) \cite{Bozza:2002zj}. The corresponding angular radius $\th_E$ and physical Einstein radius $R_E$ are determined by the reduced deflection angle ($\bar{\a} := \a\mod 2\pi$) as,
\begin{equation}
 \th_E = \f{D_{LS}}{D_S}  \bar{\a}   ~~~~ \mbox{and}~~~~R_{E} = \frac{D_{L} D_{LS}}{D_{S}} \bar{\a}  .\label{eq:gen-RE}
\end{equation}  
The analytical and numerical results of the previous section suggest that, for any given $b/b_0$ the deflection angle $\alpha$ is smaller for the WGEB model due to presence of $\d$. The 4D case can be recovered by setting $\d=0$.
Thus, for fixed $b=h/k$ (which is a {\em non-observable} in lensing data), $m$, $b_0$, and lensing distances, the WGEB deflection angle is smaller than the corresponding GEB value; consequently, the weak-lensing Einstein radius is smaller.
For weak (non-relativistic) lensing, the Einstein radius (using Eqs. (\ref{eq:gen-RE}), (\ref{eq:def-WEB2-wl}) and $R_E = b_w$), in WGEB spacetime, can be written analytically as, 
\begin{equation}
R_{0} = \left( \a_m \frac{D_{L} D_{LS}}{D_{S}} b_{0}^{m} \right)^{\f{1}{m+1}} ~~ \mbox{where}~~ \a_m = \f{2 }{ m} \le[\f{\pi}{4} + \le(\f{\sq{\pi}}{2} - \f{\sq{\pi}}{m}\ri) \f{\G(\f{m+1}{2})}{\G(\f{m}{2})} \ri] . \label{eq:RE-EB-5D}
\end{equation}
Where the subscript $0$ denotes weak lensing regime. 
The corresponding angular radius (using $b_w = D_L \th_E$) is given by
\begin{equation}
\th_{0} = \left( \a_m \frac{ D_{LS}}{D_{S}D_L^m} b_{0}^{m} \right)^{\f{1}{m+1}}. \label{eq:ang-radius-NR}
\end{equation}
Eq. (\ref{eq:ang-radius-NR}) reveals one of the distinguishing features of the wormhole parameter $m$.
The effect of the extra dimension is not explicit in Eq. (\ref{eq:ang-radius-NR}); rather, it enters implicitly through the effective impact parameter appearing in the null-geodesic equations.
For $m=2$, we get back the same result as the EB scenario \cite{Tsukamoto:2012xs} for weak lensing.

In general, the deflection angle $\alpha$ due to the presence of the lensing object can be written as $\alpha= \bar{\alpha}+2\pi n$ where $n$ is a non-negative integer, denoting the winding number of light rays around the lensing object and $\bar{\a} << 1$. 
$n\ra 0$ would imply single image formation and $n>0$ would imply relativistic (multiple) image formation by the rays passing close to the neck and looping around.
Clearly, $n$ depends on the impact parameter by the relation
\be
\bar{\a} + 2\pi n = -\pi + 2I_m(k),~~~~\mbox{where},~~ k=b_0/b_w~~,  \la{eq:alpha-ring}
\ee
and $I_m(k)$ is given by Eq. (\ref{eq:5d-da-r}) which can also be written as 
\begin{equation}
    I_m(k)= \int^{\pi/2}_0 \frac{d\t}{\left( 1- k^m\sin^m\t \right)^{\frac{m-1}{m}}}, ~~~~~~  \mbox{where},~~ u \,b= \sin \t. 
\end{equation}
Noting that $\bar{\a} = \th D_S/D_{LS}$ and $\th = b_w/D_L = b_0/kD_L$, using Eq. (\ref{eq:alpha-ring}), one arrives at the following transcendental equation for $k$,
\be
2I_m(k) - (2n+1)\pi - \f{\eta}{k} = 0,  ~~~~~~  \mbox{where},~~ \eta = \f{D_S}{D_LD_{LS}} b_0.             \la{eq:trans}           
\ee
Then the angular ring radius is given by 
\begin{eqnarray}
    \theta_n &=& \frac{b_0}{D_L k_n(m,\eta)}, \, \la{eq:ang-rad} 
\end{eqnarray}
where $k_n(m,\eta) \in (0,1)$ is a unique root of the transcendental equation (\ref{eq:trans}) for any given $m$ and $\eta$. 
The existence and uniqueness follow from the continuity and strict monotonicity of the function $F(k)=2 I_m(k)-(2n+1)\pi - \eta/k$ (a proof is provided in Appendix \ref{app:unique-proof}).
Note that $n=0$ branch approaches the weak-deflection regime, while increasing $n$ corresponds to increasingly strong-deflection relativistic images and $n\ra \infty $ approaches the critical angular scale $\th_{\infty}$.
In general, $k_n$ monotonically increases and approaches unity as $n\rightarrow \infty$ and the image angle $\theta_n$ monotonically decreases and approaches $\th_{\infty}=b_0/D_L$ for both GEB and WGEB background.
Using Eqs. (\ref{eq:ang-rad}), (\ref{eq:ang-radius-NR}), we can write
\be
 \th_0^{m+1} = \a_m \frac{D_{LS}}{D_{S} } \th_\infty^m   ; ~~~~ m=2,  4, 6, ... \la{eq:GEB_rel_Ering}
\ee
In the limit $m = 2$, we get back the 4D result \cite{Tsukamoto:2012xs}. 
\begin{equation}
    \theta_\infty = \left( \frac{4}{\pi}  \frac{D_S}{D_{LS}} \right)^{1/2} \theta_0^{3/2},
    \label{eq:WH_rel_Ering}
\end{equation}
Thus Eq. (\ref{eq:GEB_rel_Ering}) provides another generalisation of lensing formulas from the EB model to the GEB and WGEB scenarios.
Comparing with Schwarzschild black holes, the relation between the angular sizes of the Einstein ring and the relativistic Einstein ring is given by \cite{Tsukamoto:2012xs}
\begin{equation}
     \theta_{\infty} \simeq \frac{3\sqrt{3}}{4} \frac{D_S}{D_{LS}}\theta_0^{2}.
     \label{eq:BH_rel_Ering}
\end{equation}
Equations~(\ref{eq:GEB_rel_Ering}) and (\ref{eq:BH_rel_Ering}), exhibit a different functional relation between the weak Einstein-ring radius and the relativistic-ring angular scale for GEB and Schwarzschild geometries. 
One can cast this difference in terms of the following scaling relation (for the same lens-source-observer configuration):
\be
\th_{\infty} \propto \th_0^{\,\nu}.  \la{eq:theta-scaling}
\ee
where $\nu_{Sch} = 2$ and $ 1 < \nu_{GEB} \leq 3/2$.
This relation may therefore provide a model-discriminating observable when combined with independent constraints on the lensing geometry and other lens parameters. 
However, the present lensing analysis does not provide an independent determination of $\d$ from the ring-radius relation alone.
Note that $\th_n$ is simply the radius of the $n^{th}$ Einstein ring where the source is considered as a point. Therefore, magnification diverges in this limit. Below, we derive the image magnification relations for the WGEB wormhole without implementing the point-source approximation (i.e. $\b \neq 0$).

\subsection{Image magnification} 

For any axisymmetric lens, the magnification of an image is given by \cite{1992grlensing},
\be
\m  = \le|\f{\sin\b}{\sin\th} \f{d\b}{d\th}\ri|^{-1} , \la{eq:mag}
\ee
which for small angle approximation becomes
\be
\m  = \le|\f{\b}{\th} \f{d\b}{d\th}\ri|^{-1} . \la{eq:mag-small}
\ee
Thus one can compute image magnification from the lensing equation (\ref{eq:gen-lensing-eq}) which for WGEB model simplifies to
\be
\b = \th - \f{\th_0^{m+1}}{\th^m} ,  \la{eq:lensing}  
\ee
where $\th_0$ is given by Eq. (\ref{eq:ang-radius-NR}).
For WGEB scenario then one computes magnification in a straightforward manner. 
For weak-field deflection angle, we get 
\be
\m = \le|\f{\th}{\b[1 + m (\th_0/\th)^{m+1}]}\ri|. \la{eq:mag-small-WGEB}  
\ee
Eq. (\ref{eq:mag-small-WGEB} ) reduces to standard EB result for $m=2$ and $\d=0$.
Analysis of observational implications in the context of so-called microlensing by compact objects \cite{Safonova:2001vz} would be an interesting direction of future work.

To derive magnification corresponding to relativistic lensing, let us rewrite the lens equation (\ref{eq:gen-lensing-eq}) as (with $k=b_0/b_w$)
\be
\beta = \frac{b_0}{D_{L}k} - \frac{D_{LS}}{D_{S}} \le[2 I_m(k) - (2n+1)\pi \ri], \la{eq:lens-rel}
\ee
or, equivalently
\be
\f{\b}{\th} = 1 - \frac{D_{L}D_{LS}}{b_0 D_{S}}k \, \bar{\a}(k). \la{eq:lens-rel-1}
\ee
Differentiating $\b$ with respect to $k$, we get,
\be
\f{d\b}{dk} = - \f{b_0}{D_{L}k^2} - 2\frac{D_{LS}}{D_{S}}  I_m'(k) . \la{eq:derivative-mag}
\ee
The magnification then follows as
\be
\m = \le|\f{\b}{\th} \f{d\b}{dk} \f{dk}{d\th}\ri|^{-1} = \le[1 - \frac{D_{L}D_{LS}}{b_0 D_{S}}k \, \bar{\a}(k)\ri]^{-1} \le[1 + 2\f{D_L D_{LS}}{b_0 D_S} k^2 I'_m(k)\ri]^{-1} .  
\la{eq:mag-rel}
\ee
Since, $I_m'(k)$ is positive and diverges as $k \ra 1$ (see Appendix \ref{app:unique-proof}), Eq. (\ref{eq:derivative-mag}) shows the derivative $d\beta/dk$ becomes large near the critical orbit, leading to strong suppression of the individual relativistic-image magnification.
Remarkably, the magnification is obtained from the exact integral representation $I_m(k)$ rather than from the logarithmic strong-deflection expansion \cite{Bozza:2002zj}, valid for $m=2$.
Thus Eq. (\ref{eq:mag-rel}) provides yet another generalisation of lensing relations applicable to WGEB scenario. \\

\section{Observational prospects and Parameter degeneracy} \la{sec:obs} 

The weak-deflection regime is primarily useful for establishing consistency with conventional lensing theory,  where the correction terms ($\sim b_0^m/b_w^m << 1$) are negligibly small. 
The strongest sensitivity to the wormhole geometry is expected in the strong-deflection regime, where relativistic images and photon-ring observables probe $m$ and $b_w$. The parameter $\d$, however, remains degenerate with the projected impact parameter $b=h/k$ unless additional information sensitive to $\d$ is available.
To assess the observational relevance, we estimate the angular radii of the relativistic Einstein rings for representative supermassive compact objects-- Sgr A* and M87*.
Although these objects are generally interpreted as black holes, they provide useful benchmarks for estimating the angular scales predicted by the present wormhole model. \\
Since our geometry is parametrised by the throat radius $b_0$, we adopt $b_0 \sim \z\, GM/c^2 $ where $GM/c^2$ is chosen as the characteristic length, (although the ADM mass of the Ellis wormhole is zero) and $\zeta$  can take any value suitable for {\em illustration}.
Thus the angular radii, given by Eq. (\ref{eq:ang-rad}), can be written as
\be
\theta_n (m,b_0,\eta, D_L) = \frac{\zeta GM}{c^2D_L}\frac{1}{k_n(m,\eta)}. \la{eq:est-theta}
\ee
The limiting ring radius is $\th_\infty = b_0/D_L$.
For Sgr A* ($D_L  \sim 8.2 \mathrm{kpc}$, $M \sim 4.3 \times 10^{6} M_{\odot}$), $\th_\infty \sim \zeta\, 5.15 \mu\mathrm{as}$, 
and for M87* ($D_L \sim 16.8$ Mpc, $M \sim 6.5 \times 10^{9} M_{\odot}$),  $\th_\infty \sim \zeta\, 3.85 \mu\mathrm{as}$.
Thus for angular radii of relativistic images, we get $\th_n \sim \th_\infty /k_n$ with $k_n \leq 1$.
Other parameters, for example, $m$ affects $\th_n$ through the solution $k_n$ of the transcendental equation (\ref{eq:trans}).
As the winding number increases, $k_n$ approach the critical value $k_\infty=1$, and their angular positions $\th_n$ accumulate near $\th_\infty$. Since $I_m(k)$ diverges in the limit $k\ra 1^-$, even the first relativistic image is expected to satisfy $k_1\sim1^-$ (it is easy to check numerically using Eq. (\ref{eq:trans})).
Note that $\zeta$ and $k_n$ enter the observable predictions only through a degenerate combination ($\zeta/k_n$), then choosing a particular $k_n$ does not independently determine the physical scale.
For illustration purpose, we therefore choose, $\zeta =3$ and $k_n\simeq k_\infty =1$ . 
Hence our illustrative numbers scale simply as
\be
 \theta_\infty = 15.45\,\mu{\rm as} \left(\frac{\zeta}{3}\right)  \left(\frac{M}{4.3\times10^6M_\odot}\right) \left(\frac{8.2\,{\rm kpc}}{D_L}\right) ,\la{eq:es-sgr}
\ee
for Sgr A*, approximately, and
\be
\theta_\infty = 11.55\,\mu{\rm as} \left(\frac{\zeta}{3}\right) \left(\frac{M}{6.5\times10^9M_\odot}\right) \left(\frac{16.8\,{\rm Mpc}}{D_L}\right) ,\la{eq:es-m87}
\ee 
for M87*.
These angular scales are comparable to the angular resolution of the Event Horizon Telescope (EHT) and fall within the design capabilities anticipated for the next-generation EHT (ngEHT). \cite{EHT2019_I, Johnson2023ngEHT, Kardashev2013}.
Further, the angular separation between successive relativistic images is
\be
\Delta\theta_{n,n+1}
=\frac{b_0}{D_L}\left(\frac{1}{k_n}-\frac{1}{k_{n+1}}\right), \la{eq:ang-sep}
\ee
which tends to zero rapidly as $n$ increases. Although $\theta_\infty$ provides a characteristic limiting angular scale for the lens, it does not by itself determine the resolvability of individual relativistic images. Such an assessment requires the finite-order image positions and their angular separations, together with the angular resolution and sensitivity of the observing instrument.

Let us ask whether the effects of the wormhole geometry (through $m$ and $b_0$) be distinguished from the effects of the extra dimension (through $\d$), or are they observationally degenerate?
The measurements of a single lensing observable would exhibit parameter degeneracies, particularly between $b_0$ and $\d$, since both enter the deflection angles and lensing relations through the ratio $b_0/b_w$ where the effective impact parameter $b_w = b/\sq{1-\d^2}$.
The steepness parameter $m$ changes the functional dependence of the deflection angle and Einstein-ring relations, whereas $b_0$ sets the throat scale and $\d$ enters through $b_w$.
Observation of the deflection angle $\a$ gives us the ratio $b_0^m / b_w^m$. 
If $b_w$ can be reconstructed over a range of trajectories, then slope of $\log (\a_{WL})$ vs $\log (b_w)$ gives the steepness parameter-$m$ and the amplitude in Eq. (\ref{eq:WGEB-wl}) then provides the value of $b_0$.
The extra-dimensional parameter $\d$, however, remains degenerate with the underlying projected impact parameter $b=h/k$.
\\
Potentially complementary observables include flux ratios, time delays, polarization effects, frequency-dependent propagation, gravitational-wave signatures, or massive-particle dynamics; whether any of these actually breaks the $b-\d$ degeneracy is model dependent. 
Their quantitative evaluation requires solving the lens equation for specified source–lens configurations and, for realistic lens systems, modeling the lens potential, source distribution, and relevant lensing geometry \cite{Saha2024}.
A full Fisher-matrix or Bayesian parameter-estimation analysis would be an interesting direction as well \cite{Suyu2006, Balmes2013}.
A quantitative investigation along these directions lies beyond the scope of the present work and is left for future study.\\
It should be emphasized that the modified lensing properties obtained here are not, by themselves, unique to WGEB models. Similar qualitative deviations from the Ellis–Bronnikov lensing behaviour may arise in several alternative theories of gravity. The distinguishing feature of the present model is that the deviations originate from the conserved momentum associated with the extra-dimension, which rescales the effective impact parameter. 
Whether this mechanism can be observationally distinguished from other modified-gravity scenarios requires a systematic model-comparison analysis using multiple observables, which lies beyond the scope of the present work.\\

\section{Discussion}
\label{sec:discussion}

The generalized Ellis–Bronnikov (GEB) wormhole was introduced as an extension of the standard Ellis–Bronnikov (EB) spacetime with the aim of partially reducing the violation of classical energy conditions while preserving the essential wormhole structure. 
More recently, an embedding of this four-dimensional geometry into a five-dimensional warped braneworld background (WGEB) was shown to further suppress the violation of energy conditions, particularly in the presence of a decaying warp factor. 
Motivated by these developments, we have investigated deflection of null rays and gravitational lensing by both the GEB and the WGEB wormhole spacetimes.
Our primary objective was to identify potential observable signatures of the wormhole parameters, namely the throat radius $b_0$, the steepness parameter $m$, and the extra-dimensional parameter $\d$, which encodes the momentum of photons along the warped extra dimension. 
In the following, we summarise our key findings.

\begin{itemize}

\item We first analysed the null geodesics and classified photon trajectories into trapped, returning, and crossing trajectories. 
In the GEB geometry, the impact parameter $b = h/k \ge b_0$ implies the existence of returning trajectories. In the WGEB background, this condition is modified by the extra-dimensional momentum, leading to an effective impact parameter $b_w = b/\sq{1-\d^2}$. As a consequence, null rays with the same $h/k$ may experience different closest approaches depending on their momentum along the extra dimension. 
Though the geometric location of the photon sphere remains unchanged, variation in $\d$ changes the relation between the projected impact parameter $b=h/k$ and the effective impact parameter $b_w$. 
The warped extra dimension does not modify the functional form of the null-geodesic equation; rather, its effect is encoded in the effective impact parameter $b_w$. Since every lensing observable is ultimately derived from the null geodesics, the five-dimensional lensing formalism is obtained exactly from the four-dimensional one through the mapping $b\ra b_w$.

\item We obtained analytic expressions of the deflection angle in both weak- and strong-lensing regimes. In the weak-field limit, we derived a general expression for the GEB wormhole--
\begin{equation}
{}^{WGEB}\alpha_{WL} \sim   \f{2 }{ m} \le[\f{\pi}{4} + \le(\f{\sq{\pi}}{2} - \f{\sq{\pi}}{m}\ri) \f{\G(\f{m+1}{2})}{\G(\f{m}{2})} \ri]\f{b_0^{m}}{b_w^m }, \nn
\end{equation}
which reduces to the known EB result for $m=2$. This shows explicitly that the steepness parameter $m$ leaves a clear signature in the lensing behavior-- as $m$ increases, the spacetime flattens more rapidly away from the throat, resulting in weaker deflection for trajectories passing far from the throat.

\item In the strong-lensing regime, we showed that the standard EB wormhole exhibits the familiar logarithmic divergence of the deflection angle as the point of closest approach approaches the throat. In contrast, for $m>2$, the generalized geometry leads to a stronger power-law divergence as given by
\be
{}^{WGEB}\a_{SL} \approx  \sq{\f{2}{m}} B\le(\f{1}{2}-\f{1}{m}, \f{1}{2}\ri)\le(\f{b^m}{b_0^m(1-\d^2)^{m/2}}-1\ri)^{\f{1}{m}-\f{1}{2}} . \nn
\ee
Physically, this occurs due to the presence of higher order zeros, in the effective geodesic potential of photons near the throat, in higher $m$-geometries.

\item In the five-dimensional WGEB case, the analytic structure of the deflection angle remains formally similar to the four-dimensional GEB scenario, but the effective impact parameter increases due to the presence of $\d$. This reduces the deflection angle for any given $b=h/k$ and allows the divergence of the lensing angle to occur even for values of $b<b_0$, a feature absent in the GEB geometry. 

\item We subsequently investigate gravitational lensing in both weak and strong lensing limits. The Einstein radius, angular Einstein ring radii and magnification were derived analytically for the GEB/WGEB wormhole geometry. 
The steepness parameter `$m$' appears explicitly in these relations, providing a direct lensing observable. 
For a fixed projected impact parameter $b=h/k$ (which is, however, a non-observable in the lensing data), the extra-dimensional parameter produces an effective uncertainty in the inferred impact parameter and consequently in the associated image positions. This should not be interpreted as a physical broadening or displacement of the photon sphere itself.

\item A comparison with Schwarzschild lensing shows that the relation between relativistic and non-relativistic Einstein-ring radii has a different functional dependence for the GEB and Schwarzschild models. This relation may provide a model-discriminating observable when combined with independent constraints on the lensing geometry and other lens parameters.

\end{itemize}

\noindent The present work establishes the theoretical lensing framework for the GEB and WGEB wormholes.
We identify several lensing observables that differ qualitatively for different wormhole models and from their black-hole counterparts.
In the WGEB construction, $\d$ characterizes the relative extra-dimensional component of the momentum of the five-dimensional null trajectory. 
However, neither the excitation of such a bulk-propagating mode by a brane-localized source nor its subsequent detection by a brane-localized observer is modeled here. Although we found the generalization of lensing relations valid in GEB and WGEB scenarios, lensing data alone cannot constrain $\d$. 
A complete observational assessment-- including Bayesian parameter estimation, realistic instrumental uncertainties, magnifications, flux ratios and time delays-- is beyond the scope of the present work and will be addressed separately.

\section*{Acknowledgements}
Research of SJ is partially supported by the SERB, DST, Govt. of India, through a TARE fellowship grant no. TAR/2021/000354, hosted by the department of Physics, Indian Institute of Technology Kharagpur.
SG acknowledges research grant (DRIE/SMS/DRIE-05/2023-24/2352) provided by BIT Mesra.

\appendix

\section{Derivation of weak field deflection angle in GEB background} \la{app:WL-GEB}

The denominator of the integrand in Eq. \ref{eq:4d-da-x} can be written as
\be
 \f{x_0}{x} \f{\le(1 + \f{1}{x_0^m}\ri)^{1/m}}{\le(1 + \f{1}{x^m}\ri)^{1/m}} \le[ x^2 \le(1 + \f{1}{x^m}\ri)^{2/m} - x_0^2 \le(1 +  \f{1}{x_0^m}\ri)^{2/m}\ri]^{\f{1}{2}}.
\ee
Using (convergent) binomial expansions, 
\be
\le(1+ \f{1}{x_0^m}\ri)^{1/m} = \sum_{p=0}^{\infty}  \binom{\f{1}{m}}{p} \le(x_0^{-m}\ri)^{p} = 1 + \f{1}{m} \f{x_0^{-m}}{1!} + \f{1}{m}\le(\f{1}{m}-1\ri)\f{x_0^{-2m}}{2!} + \mc{O}(x_0^{-3m}) ,
\ee
\be
\le(1+ \f{1}{x^m}\ri)^{-1/m} = \sum_{q=0}^{\infty}  \binom{-\f{1}{m}}{q} \le(x^{-m}\ri)^{q} 
= 1 - \f{1}{m} x^{-m} + \f{1}{m}\le(\f{1}{m}+ 1\ri)\f{x^{-2m}}{2!} + \mc{O}(x^{-3m}) ,
\ee
\be
\mbox{where}~~~~\binom{\f{1}{m}}{k} = \f{1 (1-m)(1-2m)...(1-(k-1)m)}{k!\,m^k} ,
\ee
\be
\mbox{and}~~~~\binom{-\f{1}{m}}{k} = (-1)^k\f{1 (1+m)(1+2m)...(1+(k-1)m)}{k!\,m^k} .
\ee
The integral in Eq. \ref{eq:4d-da-x} may be expanded as
\ba 
I &= & \le[\sum_{p=0}^{\infty}  \binom{\f{1}{m}}{p} x_0^{-pm}\ri] 
 \int^{\infty}_{x_{0}} dx ~\f{x_0}{x} \le[  1 - \f{1}{m} x^{-m} + \f{1}{m}\le(\f{1}{m}+ 1\ri)\f{x^{-2m}}{2!} + \mc{O}(x^{-3m}) \ri] \nn \\
  & & \times  \le[x^2 - x_0^2 + \f{2}{m} \le(\f{x^2}{x^m} - \f{x_0^2}{x_0^m}\ri) + \f{1}{2}\f{2}{m}\le(\f{2}{m}-1\ri) \le(\f{x^2}{x^{2m}} - \f{x_0^2}{x_0^{2m}}\ri)  + \mc{O}(x^{2-3m})  \ri]^{-\f{1}{2}}   ~~ \ea
Each term in the series is integrable analytically.
Note that, only the first term, in the squared-root bracket, contributes for $m=2$. Taking ($x^2 - x_0^2$) outside the squared-root term we get,
\ba
I =  \le[\sum_{p=0}^{\infty}  \binom{\f{1}{m}}{p} \f{1}{x_0^{pm}}\ri] \int^{\infty}_{x_{0}}  \f{dx\,x_0/x}{\sq{x^2-x_0^2}} 
\le[\sum_{q=0}^{\infty}  \binom{\f{-1}{m}}{q} \f{1}{x^{q m}} \ri] 
 \le[1 + \sum_{k=1}^{\infty}  \binom{\f{2}{m}}{k} \f{ \f{x^2}{x^{km}} -  \f{x_0^2}{x_0^{km}}}{(x^2-x_0^2)} \ri]^{-\f{1}{2}}   ~~~~ \la{eq:master}
\ea
Note that the terms in the series in the last squared bracket contributes as of order $x_0^{km}$ and upto $k=m/2$ for a given $m$ thus do not contribute for $m=2$. However it is this term that makes the computation difficult for $m>2$ models.
Let us say that we are only interested upto order $x_0^{2m}$.
The above integral \ref{eq:master} can be evaluated as an expansion in $x_0^{-km}$ ($I = \sum_{k=0}^\infty I_{k}$). $I_{k}$ for first few orders  are derived below--
\ba
I_{k=0} &=& \int^{\infty}_{x_{0}} dx \f{x_0}{x \sq{x^2-x_0^2}} = \f{\pi}{2} . 
\ea
\ba
I_1 &=& \f{1}{m}  \int^{\infty}_{x_{0}} \f{dx\, x_0}{x \sq{x^2-x_0^2}} \le[x_0^{-m}-x^{-m}\ri]  + \f{1}{m} \int^{\infty}_{x_{0}} \f{dx\, x_0}{x \sq{x^2-x_0^2}} \le[\f{ x_0^{2-m} -  x^{2-m}}{2(x^2-x_0^2 )} \ri]\\
&=& \f{1}{m} \le[\f{\pi}{2} - \f{\sq{\pi}\G(\f{m+1}{2})}{m\G(\f{m}{2})}\ri] x_0^{-m} + \f{1}{m} \le[-\f{\pi}{4} + \f{\sq{\pi}\G(\f{m+1}{2})}{2\G(\f{m}{2})}\ri] x_0^{-m}\\
&=& \f{1}{m} \le[\f{\pi}{4} + \le(\f{\sq{\pi}}{2} - \f{\sq{\pi}}{m}\ri) \f{\G(\f{m+1}{2})}{\G(\f{m}{2})} \ri] x_0^{-m}    \la{eq:int1}
\ea
\ba
I_{2} &=& \int^{\infty}_{x_{0}}  \f{dx\,x_0/x}{\sq{x^2-x_0^2}} \le[ \binom{\f{1}{m}}{2} \f{1}{x_0^{2m}} + \binom{\f{-1}{m}}{2} \f{1}{x^{2 m}} + \binom{\f{1}{m}}{1} \f{1}{x_0^{m}}\binom{\f{-1}{m}}{1} \f{1}{x^{m}}\ri] \nn \\ 
 &+&  \int^{\infty}_{x_{0}}  \f{dx\,x_0/x}{\sq{x^2-x_0^2}} 
\le[ \binom{\f{1}{m}}{1} \f{1}{x_0^{m}} +  \binom{\f{-1}{m}}{1} \f{1}{x^{ m}} \ri]   \le[\binom{\f{2}{m}}{1} \f{  x_0^{2-m} - x^{2-m}}{2(x^2-x_0^2)} \ri] \nn \\   
 &+& \int^{\infty}_{x_{0}}  \f{dx\,x_0/x}{\sq{x^2-x_0^2}} 
 \le[\binom{\f{2}{m}}{2} \f{  x_0^{2-2m} - x^{2-2m}}{2(x^2-x_0^2)} \ri]  . 
\ea
Again notice that for $m=2$ only the topmost integral contributes.
For general $m$ one can evaluate the above integral in a straight forward manner to get
\ba
x_0^{2m}I_{2} &=& \le[ \f{1}{2!m}\le(\f{1}{m}-1\ri) \f{\pi}{2} + \f{1}{2!m}\le(\f{1}{m}+ 1\ri) \f{\sq{\pi}}{2} \f{\G(m+ \f{1}{2})}{\G{(m+1)}} - \f{1}{m^2} \f{\sq{\pi}}{m} \f{\G(\f{m}{2}+1)}{\G{(\f{m}{2})}}\ri] ~~ \nn \\
&& - \f{1}{m^2} \le[\f{\pi}{2} + \f{\sq{\pi} \G(m+1/2)}{\G(m)} - \f{\sq{\pi} (m+ \f{1}{2})\G(\f{m+1}{2})}{(m/2)!}\ri]  \nn \\
&& + \f{2}{2!m}\le(\f{2}{m}-1\ri) \le[\f{\pi}{4} - \f{\sq{\pi} \G(m+1/2)}{2\G(m)} \ri] .
\ea

Simplifying further, we get
\ba
I_{2} &=&  \le[-\f{3\pi}{4m^2} + \f{2\sq{\pi}}{m^2} \f{\G(\f{m+1}{2})}{\G(\f{m}{2})} - \f{\sq{\pi}}{2m^3} \f{\G(m+\f{3}{2})}{\G(m-1)} \ri] x_0^{2m} . \la{eq:int2}
\ea
It is easy to check that for $m=2$
\be
I_0 = \f{\pi}{2}, ~~~~ I_1 = \f{\pi}{8} \f{1}{x_0^2}, ~~~~  I_2 = -\f{7\pi}{128} \f{1}{x_0^4} \la{eq:I_m2}
\ee
The deflection angle is ${}^{GEB}\a = -\pi + 2I$, where $I$ is given by Eq. (\ref{eq:master}).
Eq. (\ref{eq:I_m2}) then correctly reproduces well known results for $m=2$,
\ba
{}^{EB}\a = \f{\pi}{4}\f{b_0^2}{l_0^2} - \f{7\pi}{64}\f{b_0^4}{l_0^4} + O(x_0^{-6}) 
\ea
Replacing $l_0 = \sq{b^2-b_0^2}$ above and keeping leading order terms we get
\ba
{}^{EB}\a = \f{\pi}{4}\f{b_0^2}{b^2} + \f{9\pi}{64}\f{b_0^4}{b^4}  + O\le(\f{b_0^6}{b^6}\ri)  
\ea
as reported in \cite{Nakajima:2012pu, Tsukamoto:2012xs, Tsukamoto:2012zz, Bhattacharya:2010zzb}.
Thus Eqs (\ref{eq:master}), (\ref{eq:int1}) and (\ref{eq:int2}) are the new results here.

\section{Derivation of strong field deflection angle in GEB background}\la{app:SL-GEB}

Note that for any given $m\ge 2$, the integrand in Eq. (\ref{eq:4d-da-x}) is a monotonically decreasing function of $x$ and is dominated by the values  near $x\sim x_{0+}$. Eq. (\ref{eq:4d-da-x}) can be simplified to (neglecting an overall constant factor $(1+ x_0^m)^{1/m}$),
\ba
I_{x_0 \ra 0+} &= & \int^{1}_{x_{0}} + \int^{\infty}_{1} = \int^{1}_{x_{0}} dx \f{(1+x^m)^{-1/m}}{\sq{(1+x^m)^{2/m} -(1+x_0^m)^{2/m}}} + \mbox{Finite terms} \\
&= &\int^{1}_{x_{0}} dx \f{\le[1+ \sum_{k=1}^{\infty} \binom{-1/m}{k} x^{k m} \ri]}{\sq{(1+x^m)^{2/m} -(1+x_0^m)^{2/m}}} + \mbox{Finite terms} \la{eq:approx-1}  \\
&= &\int^{1}_{x_{0}} dx \f{1}{\sq{(1+x^m)^{2/m} -(1+x_0^m)^{2/m}}} + \mbox{Finite terms} \\
&= & \sq{\f{m}{2}} \int^{1}_{x_{0}} dx \f{\le[1+ \f{m}{2}\sum_{k=2}^{\infty} \binom{2/m}{k} \f{x^{k m} - x_0^{k m}}{x^m - x_0^m} \ri]^{-1/2}}{\sq{x^m - x_0^m}} + \mbox{Finite terms} \la{eq:approx-2} \\
&= & \sq{\f{m}{2}} \int^{1}_{x_{0}} dx \f{1}{\sq{x^m - x_0^m}} + \mbox{Finite terms} \la{eq:leading}
\ea
Above integral is convergent for $m<2$. Let us analyse divergence near $x\ra 0^+$ for $m\geq 2$.
For $m=2$ we get
\be
I_{x_0 \ra 0+}^{(m=2)} \sim \int^{1}_{x_{0}} dx \f{1}{\sq{x^2 - x_0^2}} = \tanh^{-1}\sq{1-x_0^2}
\ee
which diverges exactly as $\log (2/x_0)$ thus recovering the well known logarithmic divergence.
For $m>2$, setting ($x = x_0\, t$), we get
\be
I_{x_0 \ra 0+}^{(m \geq 4)} 
\sim  x_0^{1-\f{m}{2}}  \sq{\f{m}{2}} \int^{\infty}_{1} \f{dt}{\sq{t^{m} -1}} \sim \sq{\f{1}{2m}} B\le(\f{1}{2}-\f{1}{m}, \f{1}{2}\ri)x_0^{1-\f{m}{2}}. 
\ee
Where $B(p,q)$ is the beta function.
Let us further look at the sub-leading term (of order $x^{m/2}$) within the ``finite terms'' in Eq. (\ref{eq:leading}) (for validation of Fig. \ref{fig:int-approx}). Using Equations  (\ref{eq:approx-1}) and (\ref{eq:approx-2}), we get 
\be
I_{x_0 \ra 0+}^{sub} = \eta_m \sq{\f{m}{2}} \int^{1}_{x_{0}} dx \f{x^m}{\sq{x^m - x_0^m}}, ~~~~ \mbox{where}~~~~ \eta_m = \binom{-\f{1}{m}}{1}- \f{m}{4} \binom{\f{2}{m}}{2}  .\la{eq:sub-m}
\ee
Note that $\eta_m =0$ for $m=6$,  $\eta_m <0$ for $m<6$ and $\eta_m >0$ for $m>6$. 
Thus, for $m=6$, the sub-sub-leading term (of order $x^{3m/2}$) contributes as
\be
I_{x_0 \ra 0+}^{sub-sub;m=6} = \eta_6 \sq{\f{m}{2}} \int^{1}_{x_{0}} dx \f{x^{2m}}{\sq{x^m - x_0^m}} , ~~~~ \mbox{where}~~ \eta_6 = \binom{-\f{1}{m}}{2}- \f{m}{4} \binom{\f{2}{m}}{3} = \f{1}{216} . \la{eq:sub-6}
\ee
Thus Equations (\ref{eq:sub-m}) and (\ref{eq:sub-6}) correctly explains the change in sign of $(\f{I_{approx}}{I_{numeric}} -1)$ for cases $m=4$ and $m>4$ as depicted in Fig. \ref{fig:int-approx}.

\section{ Non-uniformity of the joint limit $m \ra 2^+, x_0 \ra 0^+$} \la{app:limit-m-2}
As given by Eq. (\ref{eq:SL-approx-EB}), for exactly $m=2$, we have $ \a_{\rm SL} (m=2, x_0 \ra 0^+) \sim -2\ln x_0-\pi $.
For $m>2$, the asymptotic expansion, Eq.  (\ref{eq:SL-approx-GEB}) implies,
\be
\a_{\rm SL}  (m,x_0 \ra 0^+) \sim A_m x_0^{-(m-2)/2}, ~~~~ \mbox{where}~~~~ A_m = \sq{\f{2}{m}} B\le(\f{1}{2}-\f{1}{m}, \f{1}{2}\ri) .
\ee
Let us write $ \ep=m-2$. Then in the limit $ \ep \ra 0^+$, one can write, 
\ba 
x_0^{-\ep/2} &=& \exp[-\ep (\ln x_0)/2] = 1-\frac{\ep}{2}\ln x_0 +O(\ep^2), \\
\mbox{and}~~~~  A_m & = & \sq{\f{2}{2+\ep}} B\le(\f{\ep}{4}, \f{1}{2}\ri) =   \frac{4}{\ep} +\text{finite terms}.
\ea
Therefore, in the limit $m\ra 2+$, we have, 
\be
\lim_{m\ra2+} \a_{\rm SL} \simeq \frac{4}{\ep} - 2\ln x_0 + \cdots
\ee
The divergent $4/\ep$ term is an $x_0$-independent contribution arising from the non-uniform asymptotic expansion and therefore cannot be identified with the $m=2$ logarithmic divergence. 
While the $x_0$-dependent term contains $-2\ln x_0$, which is precisely the logarithmic structure emerging at $m=2$.
Thus the $m>2$ power-law asymptotic does not possess a uniform limit at $m=2$ and the $m=2$ logarithmic behaviour must be treated separately.

\section{Angular radius and image position} \la{app:ang}
One can calculate the image position by using the following formulae, 
\begin{equation}
\beta = \theta - \frac{\theta_{E}^{3}}{\theta^2}, ~~\text{for} \left( \theta > 0 \right) \label{eq:image-position-EB1}
\end{equation}
\begin{equation}
\mbox{and}~~~~\beta = \theta + \frac{\theta_{E}^{3}}{\theta^2}, ~~\text{for} \left( \theta < 0 \right), \label{eq:image-position-EB2}
\end{equation}
where $\theta = b/D_{L} \approx l_{0}/D_{L}$ and $\theta_{E} = R_{E}/D_{L}$ is the angular radius of Einstein ring. We can reduce the above equations into simple cubic equations, given below, via defining new parameters as follows, $\hat{\beta} = \beta/\theta_{E}$ and $\hat{\theta} = \theta/\theta_{E}$.
\begin{equation}
\hat{\theta}^{3} - \hat{\theta}^{2} \hat{\beta} - 1 = 0, ~~\text{for} \left( \hat{\theta} > 0 \right)\label{eq:cubic-eq-1}
\end{equation}
\begin{equation}
\mbox{and}~~~~ \hat{\theta}^{3} - \hat{\theta}^{2} \hat{\beta} + 1 = 0, ~~\text{for} \left( \hat{\theta} < 0 \right) .\label{eq:cubic-eq-2}
\end{equation}

Since, the discriminant of Eqs. (\ref{eq:cubic-eq-1}) is $-4\hat{\beta}^{3} - 27 < 0$, therefore, Eqs. (\ref{eq:cubic-eq-1}) would have two complex conjugate solutions and one real solution ( using Cardano's formula), given by,
\begin{equation}
\hat{\theta} = \frac{\hat{\beta}}{3} + U_{1+} + U_{1-} \label{eq:theta-solution-1},
\end{equation}
\begin{equation}
\mbox{where}~~~~ U_{1\pm} = \left( \frac{\hat{\beta}^{3}}{27} + \frac{1}{2} \pm \sqrt{\frac{1}{4} \left( 1 + \frac{2\hat{\beta}^{3}}{27} \right)^{2} - \frac{\hat{\beta}^{6}}{27^{2}}} \right)^{1/3} \label{eq:U1}.
\end{equation}
The real positive solution implies the physical images of the source. Similarly, the discriminant of Eq. (\ref{eq:cubic-eq-2}) is $4\hat{\beta}^{3} - 27$, and so, if $\hat{\beta} < \left( 27/4 \right)^{1/3}$ then the corresponding equation has one real solution:
\begin{equation}
\hat{\theta} = \frac{\hat{\beta}}{3} + U_{2+} + U_{2-} \label{eq:theta-solution-2}
\end{equation}
\begin{equation}
\mbox{where}~~~~U_{2 \pm} = \left( \frac{\hat{\beta}^{3}}{27}  - \frac{1}{2} \pm \sqrt{\frac{1}{4} \left( 1 - \frac{2\hat{\beta}^{3}}{27} \right)^{2} - \frac{\hat{\beta}^{6}}{27^{2}}}\right)^{1/3} \label{eq:U2}.
\end{equation} 
The image will form inside the Einstein ring corresponding to solution given by Eq. (\ref{eq:theta-solution-2}). 
The difference in image positions for 4D and 5D models appears through $\th_E$ or the Einstein radius $R_E$ which depends on the parameter $\delta$,  as discussed above. 


\section{Proof of existence and uniqueness of the solution of the transcendental equation defining the relativistic Einstein rings } \la{app:unique-proof}

\be
\mbox{Let }~~~~F(k) = 2I_m(k) - (2n+1)\pi - \f{\eta}{k}   , ~~~~ n \in N, \eta>0, k \in (0,1)          
\ee
\begin{equation}
\mbox{where}~~~~    I_m(k)= \int^{\pi/2}_0 \frac{d\t}{\left( 1- k^m\sin^m\t \right)^{\frac{m-1}{m}}}, ~~~~m \geq 2 .  \la{eq:Im}
\end{equation}
Then the transcendental equation that we need to solve is $F(k)=0$.
Since $0<k<1$, $I_m(k)$ and therefore $F(k)$ is a continuous function.
In the $k\ra 0$ limit, $I_m \ra \pi/2$ and thus $F(k) \ra - \infty$. Whereas in the $k\ra 1$ limit, $I_m$ diverges (logarithmically for $m=2$ and as a power law for $m>2$ as shown in Appendix \ref{app:SL-GEB} above) and thus $F(k) \ra + \infty$. 
Since $F(k)$ is continuous and $F(0^+)=-\infty, F(1^-)=+\infty$, by the Intermediate Value Theorem \cite{Rudin1976}, there {\em exists} at least one solution $k_n \in (0,1)$ to the equation $F(k)=0$.\\
To prove the {\em uniqueness}, $F(k)$ needs to be monotonic, i.e. $F'(k) = 2 I_m'(k) + \eta/k^2 > 0 \, \forall \, k \in (0,1)$ since a strictly increasing function can cross zero only once.
Differentiating Eq. (\ref{eq:Im}), we get
\begin{equation}
    I_m'(k)= (m-1)k^{m-1} \int^{\pi/2}_0 \frac{\sin^m\t \,d\t}{\left( 1- k^m\sin^m\t \right)^{\frac{2m-1}{m}}} .  \la{eq:I'm}
\end{equation}
Since the integrand is positive everywhere $\t \in [0, \pi/2] \implies I_m' >0 \implies F'(k) >0$. 
This proves the existence and uniqueness of the solution of the transcendental equation $F(k)=0$.

\section*{Bibliography} 
\bibliographystyle{unsrt}
\bibliography{References}


\end{document}